\documentclass[twoside,12pt]{article}%
\usepackage{amsmath}
\usepackage{epsfig}
\usepackage{amsfonts}
\usepackage{amssymb}
\usepackage{appendix}
\usepackage{graphicx}
\usepackage{arydshln}
\usepackage{multirow}%
\providecommand{\U}[1]{\protect\rule{.1in}{.1in}}

\def\ss{\mbox{\boldmath $\sigma$}}
\newcommand{\be}{\begin{equation}}
\newcommand{\ee}{\end{equation}}
\newcommand{\bea}{\begin{eqnarray}}
\newcommand{\eea}{\end{eqnarray}}

\begin{document}

\title{A simple alternative to the relativistic Breit-Wigner distribution}
\author{Francesco Giacosa$^{(1,2)}$, Anna
Okopi\'{n}ska$^{(1)}$, Vanamali Shastry$^{(1)}$\\$^{(1)}$\emph{Institute of Physics, Jan Kochanowski University, }\\\emph{ul.  Uniwersytecka  7,  P-25-406  Kielce,  Poland}\\$^{(2)}$\emph{Institute for Theoretical Physics }\\\emph{Johann Wolfgang Goethe - University, Max von Laue--Str. 1}\\\emph{D-60438 Frankfurt, Germany} }
\maketitle

\begin{abstract}
First, we discuss the conditions under which the non-relativistic and relativistic
types of the Breit-Wigner energy distributions are obtained. Then, upon insisting on
the correct normalization of the energy distribution, we introduce a
Flatt\'{e}-like relativistic distribution -denominated as Sill distribution- that (i) contains left-threshold
effects, (ii) is properly normalized for any decay width, (iii) can be obtained as an appropriate
limit in which the decay width is a constant, (iv) is easily generalized to
the multi-channel case (v) as well as to a convoluted form in case of a decay chain and - last but not least - (vi) is simple to deal with.
We compare the Sill distribution to spectral functions derived within specific QFT models and
show that it fairs well in concrete examples that involve a fit to experimental data for the $\rho$, $a_1(1260)$, and $K^*(982)$ mesons as well as the $\Delta(1232)$ baryon. We also present a study of the $f_2(1270)$ which has more than one possible decay channels. Finally, we discuss the limitations of the Sill distribution using the $a_0(980)$-$a_0(1450)$ and the $K_0^\ast(700)$-$K_0^\ast(1430)$ resonances as examples.

\end{abstract}

\section{Introduction}

The parametrization of the energy distribution (or spectral functions) of
resonances is an important topic of both Quantum Mechanics (QM) and Quantum
Field Theory (QFT). Namely, new resonances are often obtained by fits involving
 the line-shape of unstable states, see e.g. the Particle
Data Group (PDG) \cite{pdg}, and the activities in various ongoing and planned
experiments \cite{gluex,gutschegluex,Ablikim:2016qzw,Ablikim:2020cyd,compass,bes,lhcb,belle2,panda}.

The very well known Breit-Wigner (BW) distribution \cite{ww,breit}:
\begin{equation}
d_{S}(E)=d_{S}^{\text{BW}}(E)=\frac{\Gamma}{2\pi}\frac{1}{(E-M)^{2}+\frac{\Gamma^{2}%
}{4}}\text{  ,}%
\end{equation}
where $M$ is the energy (or mass) of
the unstable state and $\Gamma$ the decay width (the lifetime being simply
$1/\Gamma$ in natural units), has been widely used in countless applications.
Quite remarkably, the BW distribution has no
low-energy threshold and the corresponding decay law is exactly exponential.
It is correctly normalized, $\int_{-\infty}^{+\infty}\mathrm{dE}d_{S}^{\text{BW}}(E)=1$,
allowing for the standard interpretation of $d_{S}(E)\mathrm{dE}$ as the
probability that the unstable state possesses an energy (or mass) between $E$
and $E+\mathrm{dE.}$ It may be regarded as a very good approximation in most
cases, especially in the context of QM when $\Gamma\ll M.$ In general,
$\Gamma$ is itself a function of energy and is interpreted as the imaginary
part of the self-energy of the one-particle unstable state; the BW function
naturally arises as the limiting case in which the coupling of the unstable
state to the final decay products is energy independent.

In the relativistic case, the Breit-Wigner formula (rBW) for an unstable state $S$ reads \cite{pdg,kycia,salam} (see later on for details):
\begin{equation}
d_{S}^{\text{rBW}}(E)=\frac{2E}{\pi}\frac{M\Gamma}{(E^{2}-M^{2})^{2}%
+(M\Gamma)^{2}}\theta(E)\text{ ,}%
\label{rBW}
\end{equation}
where, for relativistic particles, the variable $E$ represents the particle running mass 
and must be non-negative. 
The
normalization of the spectral function, necessary for
its probabilistic energy interpretation, is not fulfilled since $\int_{0}^{\infty}dEd_{S}^{\text{rBW}}(E)<1$.
The loss of normalization, even if small for narrow states, is due to the fact the limiting process that works in the QM case does
not apply in the relativistic case.
Technically speaking, when the imaginary part of the self-energy  is a constant for $E\geq0$, the real part of it 
is not a simple constant that can be subtracted. 
 

In this work we aim to show that a rather simple spectral
function can be obtained as a natural limit of the relativistic context. First,
we introduce the threshold \ $E_{th}$ associated with a certain unstable state 
(for a two-body decay it is simply $E_{th}=m_{1}+m_{2},$ where $m_{1}$ and
$m_{2}$ are the rest masses of the produced particle). We then introduce a
Flatt\'{e}-like spectral function \cite{flatte,baru} that we call 
\textquotedblleft Sill\textquotedblright 
(Sill stands for threshold and for the form of the function close to it):
\begin{equation}
d_{S}^{\text{Sill}}(E)=\frac{2E}{\pi}\frac{\sqrt{E^{2}-E_{th}^{2}}%
\tilde{\Gamma}}{(E^{2}-M^{2})^{2}+(\sqrt{E^{2}-E_{th}^{2}}\tilde{\Gamma})^{2}%
}\theta(E-E_{th}) \text{ ,} \label{Sill}%
\end{equation}
where the rescaled width $\tilde{\Gamma}$ is given by $\tilde{\Gamma}=\Gamma
M/\sqrt{M^{2}-E_{th}^{2}}$ (for a nonrelativistic Flatt\'{e}-like parameterization, see Refs. \cite{Bogdanova:1991zz,Gong:2016hlt,Baru:2010ww}).

The function $d_{S}^{\text{Sill}}(E)$ is always normalized to $1$ ($\int
_{E_{th}}^{\infty}\mathrm{dE}d_{S}^{\text{Sill}}(E)=1$) for any $M$ and $\Gamma$. The form $d_{S}^{\text{Sill}}(E)$ 
is very similar to the relativistic BW distribution(s) when $\Gamma\ll M,$
but is utterly different when $\Gamma/M$ is sizable. Moreover, the constraints
of complex analysis are correctly taken into account (details of its
derivation and properties are in the paper): the function can be seen as the
proper limit of a distribution in which the decay width is a constant (the
real part of the self-energy vanishes). Quite importantly, $d_{S}%
^{\text{Sill}}(E)$ is a continuous function for any value of $E$. 

We therefore argue that a fit to data using the function 
in\ Eq. (\ref{Sill})
may be advantageous. If BW or rBW work well, then the Sill also does: it simply
delivers similar results. Yet, whenever a threshold is close enough to the
mass of the unstable state, $d_{S}^{\text{Sill}}(E)$ fairs better, as we
shall show with some concrete examples.

Indeed, Eq. (\ref{Sill}) is closely reminiscent of the Flatt\'{e} distribution \cite{flatte} employed for the resonance $a_{0}(980)$ meson decaying into
$two$ channels, $\bar{K}K$ and $\pi\eta$, where however typically only the $K\bar{K}$
channel is treated as energy dependent, while the $\pi\eta$ decay channel is
treated as a constant, see e.g. Ref. \cite{baru}.  In the context of the resonance $a_{0}(980)$, the $\bar{K}K$
channel is very close to the nominal mass, hence the peculiar shape of the
spectral function can be described. 

Our claim is that Eq. (\ref{Sill}) -as well as multi-channel and /or convoluted extensions of it- can be also used independently on the location of threshold(s).  
In addition, the Sill distribution can be applied as it stands to the case in which the decay product consists of particles of different masses without any loss of normalization of the spectral function\footnote{As it shall be clear later on, in the case of decay products with different masses, the decay width proportional to the three-momentum $k$ (which is in some cases used in the context of Flatt\'{e} approach) breaks the normalization of the spectral function.}. Summarizing, it may be regarded as a useful phenomenological approximation able to capture in first approximation 
some essential (but of course not all) features needed to properly describe various physical resonances. In this respect, it must be seen as a valid alternative to the rBW function because it contains threshold effect(s) and is mathematically consistent. Surely, it is not an alternative to a detailed study of a given resonance, in which the proper coupling to the decay products is taken into account. Yet, it can be used for both standard and non-standard mesons (such as the $a_0(980)$ mentioned above), since it is not dependent on the internal structure of hadrons

\begin{figure}
    \centering
    \includegraphics[scale=0.23]{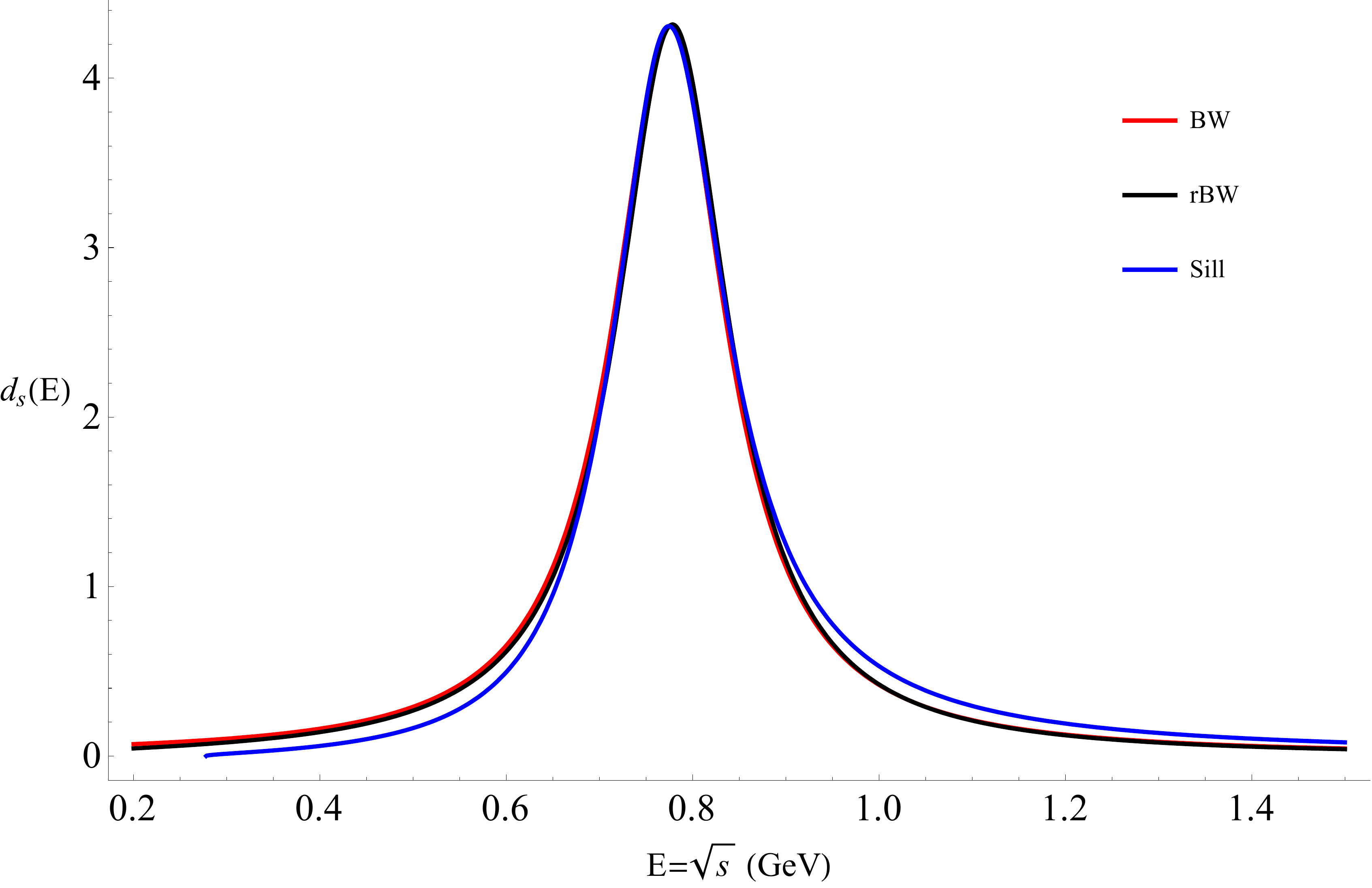}~\includegraphics[scale=0.23]{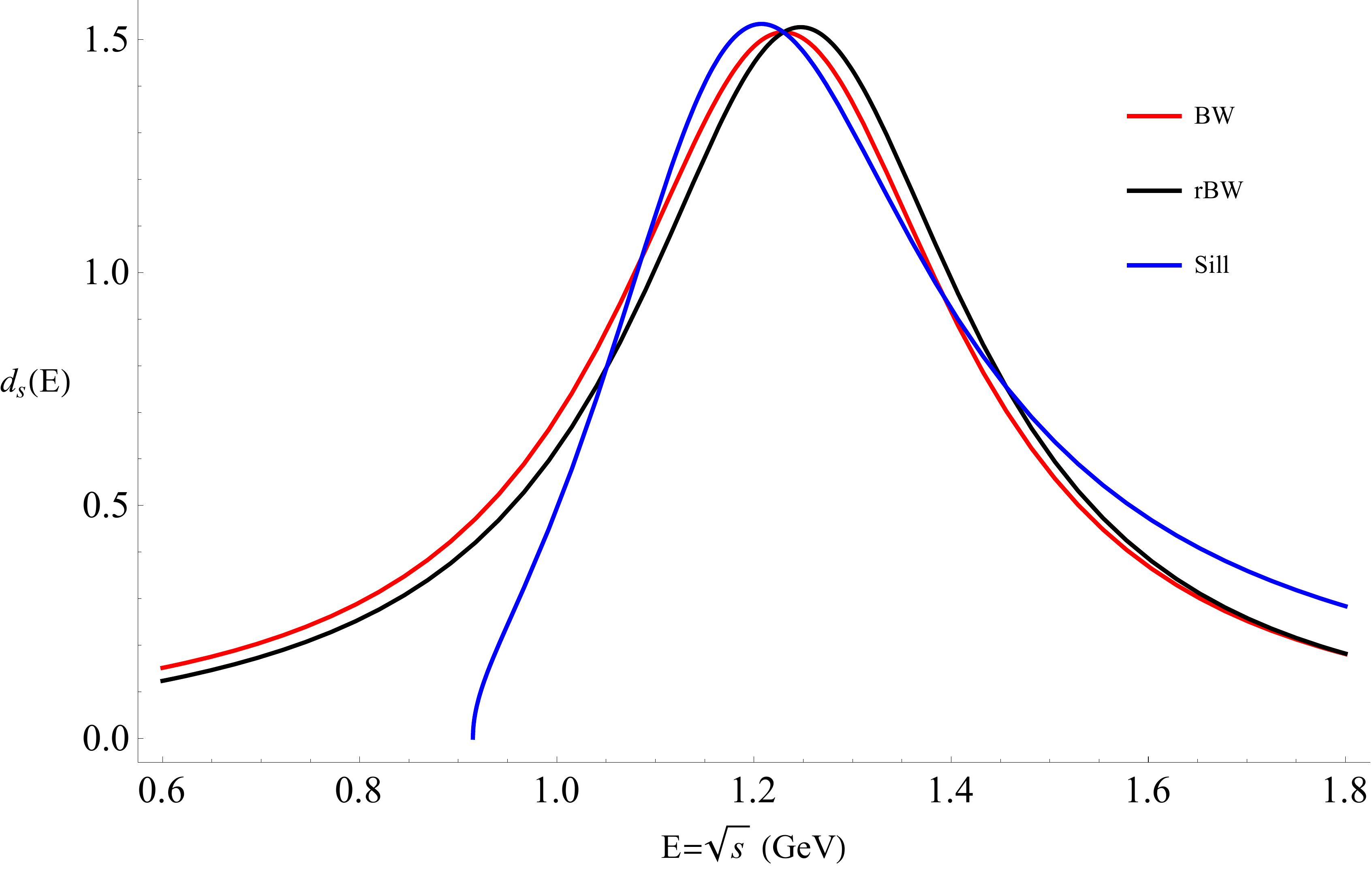}
    \caption{Illustrative comparison of the four distributions discussed in the paper. Left panel, peak far away from the threshold ($\rho(770),~M=0.775\rm{~GeV},~\Gamma=0.1478\rm{~GeV}, ~\rm{and}~E_{th}=2m_{\pi}$); right panel, peak near the threshold ($a_1(1260), ~M=1.230\rm{~GeV},~\Gamma=0.5\rm{~GeV}, ~\rm{and}~E_{th}=m_\rho+m_\pi$).}
    \label{comp}
\end{figure}

In Fig. \ref{comp} we compare the plots of the BW, rBW, and Sill distributions for the $\rho$ and $a_1(1260)$ mesons using the parameters of the PDG. This picture serves a first illustration of the differences of these functions, which are evident for broad resonances (right plot). In the following, for these two resonances as well as for the narrow $K^*(892)$ we shall perform a fit to data. We shall see that the Sill fairs better than BW and rBW for the $\rho$ and especially for the $a_1(1260)$, for which the threshold effects are important, but there is basically no difference for the narrow $K^*(892)$, since then Sill reduces to rBW in the small-width limit.

The paper is organized as follows: in Sec. 2 we discuss the non-relativistic case by using the propagator formalism and derive the standard BW spectral function; in Sec. 3 we introduce the relativistic propagator, the standard relativistic approximations as well as the decay of a scalar QFT (note, Secs. 2 and 3 contain known results which are useful for the comparison with the Sill distribution); in Sec. 4 we introduce the Sill distribution, compare it to the results of Sec. 3 and extend it to the multi-channel and convoluted cases; in Sec. 5 we discuss various numerical examples, that include fit of the BW, rBW, and the Sill to experimental data for the resonances  $\rho$, $a_1(1260)$, and $K^*(982)$ mesons and the $\Delta(1232)$ baryon. Moreover, we shall study the $f_2(1270)$ as an example of multi-channel decays. In Sec. 6, we present the $a_0(980)$ and $k$ as examples of non-conventional states, and present the limitations of our approach. In Sec. 7, we discuss the domain fo applicability of the Sill and give a empirical condition when the use of Sill will be favorable. Finally, in Sec. 8 we present our conclusions. 

\section{Brief review of the non-relativistic case}

In this section we recall the derivation of the non-relativistic BW function.
To this end, we start from the propagator of an unstable quantum state
$\left\vert S\right\rangle $ as function of the energy $E$:%
\begin{equation}
G_{S}(E)=\frac{1}{E-M+\Pi(E)+i\varepsilon}\text{ ,}\label{nrprop}%
\end{equation}
where the self-energy function $\Pi(E)$ for complex $E$ can be expressed via
the dispersion relation 
\begin{equation}
\Pi(E)=-\int_{E_{th}}^{\infty}\frac{1}{\pi}\frac{\operatorname{Im}%
\Pi(E^{\prime})}{E-E^{\prime}+i\varepsilon}dE^{\prime}\text{,}%
\end{equation}
with $E_{th}$ being the lowest energy threshold for the decay of the unstable
state. (For a well defined theoretical setup in which the above equations can
be defined, we refer to the Lee (or Friedrichs) model, see Ref.
\cite{lee,scully,facchispont,zhou,leerev,duecan,ceci} and refs. therein).

The function
\begin{equation}
\Gamma(E)=2\operatorname{Im}\Pi(E)
\end{equation}
is the so-called decay width function, whose specific form depends on the
details of the interaction. A possible definition of the mass of the unstable
state $S$ is obtained via%
\begin{equation}
M_{r}-M+\operatorname{Re}\Pi(M_{r})=0\text{ ,}%
\end{equation}
which shows that quantum fluctuations may shift the bare mass $M$. Of course, with a
suitable subtraction, one can impose that $\operatorname{Re}\Pi(M)=0,$ thus
$M_{r}=M.$ We shall adopt this choice, thus $M$ can be regarded as the nominal energy/mass of the state. The on-shell decay width is obtained by
setting $\Gamma=\Gamma(M).$ Alternatively, a commonly used definition is the
pole energy/mass $z_{pole}$ obtained by the equation
\begin{equation}
z_{pole}-M+\Pi_{II}(z_{pole})=0\text{ ,}%
\end{equation}
where $II$ refers to the second Riemann sheet. Then:
\begin{equation}
z_{pole}=M_{pole}-i\frac{\Gamma_{pole}}{2}\text{ .}%
\end{equation}
If more poles are present, typically the one(s) closest to the real axis is (are) the
most important, but exceptions are possible.
The position of the pole, being independent of the particular
reaction needed to form the unstable state, is regarded as a stable feature of
a given resonance.

The spectral function is obtained as the imaginary part of the propagator
\begin{equation}
d_{S}(E)=-\frac{1}{\pi}\operatorname{Im}[G_{S}(E)]=\frac{1}{\pi}%
\frac{\operatorname{Im}\Pi(E)}{(E-M+\operatorname{Re}\Pi(E))^{2}%
+(\operatorname{Im}\Pi(E))^{2}}\text{ ,}%
\end{equation}
or conversely%
\begin{equation}
G_{S}(E)=\int_{E_{th}}^{\infty}\frac{d_{S}(E^{\prime})}{E-E^{\prime
}+i\varepsilon}\mathrm{dE}^{\prime}
\end{equation}
in agreement with the interpretation of $\mathrm{dE}~\!d_{S}(E)$ as a mass/energy
probability between $E$ and $E+\mathrm{dE}$ \cite{salam,lupo}.
Namely, the full propagator has
been expressed as the `sum' of free propagtors with mass/energy $E^{\prime},$ each
one weighted by $d_{S}(E^{\prime}).$ 

The function $d_{S}(E)$ is correctly normalized to unity as long as
$\lim_{E\rightarrow\infty}\left\vert \Pi(E)\right\vert /E=0$ (see Refs.
\cite{leerev,lupofermions}):%
\begin{equation}
\int_{E_{th}}^{\infty}\mathrm{dE}~\!d_{S}(E)=1\text{ .}%
\end{equation}

In fact, one may consider the state $\left\vert S\right\rangle $ as the
eigenstate of the unperturbed Hamiltonian $H_{0}$ which fulfills the
normalization condition $\left\langle S|S\right\rangle =1.$ The full set of
eigenstates of the Hamiltonian $H$ reads $\{\left\vert E\right\rangle \}$ with
$H\left\vert E\right\rangle =E\left\vert E\right\rangle $ and $E\geq E_{th}$.
Expressing $\left\vert S\right\rangle $ in terms of $\left\vert E\right\rangle
$ implies%
\begin{equation}
\left\vert S\right\rangle =\int_{E_{th}}^{\infty}\mathrm{dE}~\!a_{S}(E)\left\vert
E\right\rangle \text{ .}%
\end{equation}
The quantity $d_{S}(E)=\left\vert a_{S}(E)\right\vert ^{2}$ is the `spectral
function' which is evaluated in this work as the imaginary part of the
propagator (see also Ref. \cite{duecan} for a more detailed discussion of
these relations). It naturally follows that%
\begin{equation}
1=\left\langle S|S\right\rangle =\int_{E_{th}}^{\infty}\mathrm{dE}d_{S}(E)\text{ .}%
\end{equation}

The survival probability amplitude as function of time reads:
\begin{equation}
a_{S}(t)=\frac{i}{2\pi}\int_{-\infty}^{+\infty}\mathrm{dE}G_{S}(E)e^{-iEt}%
=\int_{E_{th}}^{+\infty}\mathrm{dE}d_{S}(E)e^{-iEt}\text{.}%
\end{equation}
In general, the decay is not exponential, see Refs. \cite{facchispont,nonexpqft,kelkar,ghirardi}
for a description of this issue. Quite interestingly, also in presence of
deviations from the exponential decay, the pole contribution implies that
\begin{equation}
a_{S}(t)=Ze^{-iz_{pole}t}+...\text{,}%
\end{equation}
where the exponential contribution turns out to be dominant for intermediate times.

The Breit-Wigner limit is obtained in two steps. First, let us consider
$\Gamma(E)=\Gamma=const$ and introduce a maximal energy $\Lambda$:%
\begin{equation}
\Pi(E)=-\int_{E_{th}}^{+\Lambda}\frac{\mathrm{dE}^{\prime}}{2\pi}\frac{\Gamma
}{E-E^{\prime}+i\varepsilon}=\frac{\Gamma}{2\pi}\left[  \ln(E^{\prime
}-E)\right]  _{E_{th}}^{\Lambda}=\frac{\Gamma}{2\pi}\ln\frac{\Lambda-E}%
{E_{th}-E}\text{ .}%
\end{equation}
It follows that%
\begin{align}
\operatorname{Im}\Pi(E)  &  =\left\{
\begin{array}
[c]{c}%
\frac{\Gamma}{2}\text{ for }E\in(E_{th},\Lambda)\\
0\text{ otherwise}%
\end{array}
\right.  \text{ ;}\\
\operatorname{Re}\Pi(E)  &  =\frac{\Gamma}{2\pi}\ln\left\vert \frac{\Lambda
-E}{E_{th}-E}\right\vert \text{ .}%
\end{align}
Next, one considers the limit $E_{th}\rightarrow-\infty$ and $\Lambda
\rightarrow\infty$ together with the constraint $E_{th}/\Lambda=-1.$ Clearly,
$\operatorname{Im}\Pi(E)=\Gamma/2$ for each $E.$ On the other hand:
\begin{equation}
\operatorname{Re}\Pi(E)=\frac{\Gamma}{2\pi}\ln\left\vert 1\right\vert =0\text{
;}%
\end{equation}
(another choice for the $E_{th}/\Lambda$ ratio could be re-absorbed in the
definition of $M.$) Hence, the Breit-Wigner propagator reads%
\begin{equation}
G_{S}^{\text{BW}}(E)=\frac{1}{E-M+i\Gamma/2+i\varepsilon} \text{ ,}%
\end{equation}
whose pole is simply given by
\begin{equation}
z_{pole}^{\text{BW}}=M-i\Gamma/2 \text{ .}
\end{equation}
The spectral function in this case takes the very well known Breit-Wigner
form
\begin{equation}
d_{S}(E)=d_{S}^{\text{BW}}(E)=\frac{\Gamma}{2\pi}\frac{1}{(E-M)^{2}%
+\frac{\Gamma^{2}}{4}} \text{ .}
\end{equation}
The survival probability amplitude is in this case \textit{exactly} an
exponential:%
\begin{equation}
a_{S}(t)=a_{S}^{\text{BW}}(t)=e^{-iMt-\Gamma t/2}.
\end{equation}
The normalization%

\begin{equation}
\int_{-\infty}^{+\infty}d_{S}^{\text{BW}}(E)dE=1
\end{equation}
is clearly fulfilled.

Yet, the Breit-Wigner distribution can be only considered as a useful but unphysical limit, 
the most clear drawback
being the lack of the left energy threshold (the energy is unbounded from
below).\ Yet, it is very good close to the peak for sufficiently narrow (alias long-lived) quantum states.

\section{Brief review of the relativistic case}

In this section we present how the previous formalism can be generalized to
the relativistic case. We shall describe the series of approximations
leading to the relativistic BW distribution. Moreover, we shall also discuss
a scalar QFT as a notable example.

\subsection{Propagator and spectral function}

The relativistic case can be easily obtained by replacing the variable $E$
with the variable $s=E^{2}$ into Eq. (\ref{nrprop}), hence the propagator
(as function of $s$) takes the form (see e.g. \cite{peskin,zee})
\begin{equation}
G_{S}(s)=\frac{1}{s-M^{2}+\Pi(s)+i\varepsilon}\text{ .}%
\end{equation}
It is clear from the very beginning that $s\geq0,$ hence in a
relativistic framework a left threshold is built in. The
self-energy reads \cite{boglione}
\begin{equation}
\Pi(s)=-\int_{s_{th}}^{\infty}\frac{1}{\pi}\frac{\operatorname{Im}%
\Pi(s^{\prime})}{s-s^{\prime}+i\varepsilon}ds^{\prime}\text{,}%
\end{equation}
with $s_{th}=E_{th}^{2}\geq0$ is the low energy threshold. In this case, the
imaginary part takes the form
\begin{equation}
\operatorname{Im}\Pi(s)=\sqrt{s}\Gamma(s)\text{ .}%
\end{equation}
The mass is obtained by
\begin{equation}
M_{r}^{2}-M^{2}+\operatorname{Re}\Pi(M_{r}^{2})=0 \text{ .}
\end{equation}
Also in this case, a suitable subtraction can be made to assure that
$\operatorname{Re}\Pi(M^{2})=0,$ and hence, $M_{r}=M.$ The corresponding on-shell decay
width is then $\Gamma=\Gamma(M)$, see below.

Alternatively, the pole position $s_{pole}$ is such that
\begin{equation}
s_{pole}-M^{2}+\Pi_{II}(s_{pole})=0
\end{equation}
and leads to the pole mass and decay width defined as:
\begin{equation}
\sqrt{s_{pole}}=M_{pole}-i\frac{\Gamma_{pole}}{2}\text{ .}%
\end{equation}

The spectral function as function of $s$ is obtained as the imaginary part of
the propagator
\begin{equation}
d_{S}(s)=-\frac{1}{\pi}\operatorname{Im}[G_{S}(s)]=\frac{1}{\pi}%
\frac{\operatorname{Im}\Pi(s)}{(s-M^{2}+\operatorname{Re}\Pi(s))^{2}%
+(\operatorname{Im}\Pi(s))^{2}}\text{ ,}%
\end{equation}
then%
\begin{equation}
G_{S}(s)=\int_{s_{th}}^{\infty}\frac{d_{S}(s^{\prime})}{s-s^{\prime
}+i\varepsilon}\mathrm{ds}^{\prime}%
\end{equation}
is the K\"{a}llen-Lehmann representation in QFT, e.g. Ref. \cite{peskin}.

As function of the energy $E$ one obtains\footnote{Note, this is a slightly
incorrect notation. More correctly, one should write $d_{S}(s)ds=\tilde{d}%
_{S}(E)dE,$ thus naming the spectral function as function of $E$ differently.
Yet, in order not to overburden the notation, we replace $\tilde{d}%
_{S}(E)\rightarrow d_{S}(E).$}%
\begin{equation}
d_{S}(E)=\frac{2E}{\pi}\frac{\operatorname{Im}\Pi(E^{2})}{(E^{2}%
-M^{2}+\operatorname{Re}\Pi(E^{2}))^{2}+(\operatorname{Im}\Pi(E^{2}))^{2}} \text{ .}%
\end{equation}
Just as in the non-relativstic QM case%
\begin{equation}
\int_{s_{th}}^{\infty}d_{S}(s)ds=1\text{ , }\int_{E_{th}}^{\infty}d_{S}(E)dE=1
\end{equation}
as long as $\lim_{s\rightarrow\infty}\left\vert \Pi(s)\right\vert /s=0$.

In general, the function $\Pi(s=E^{2})$ depends on the employed model.
The imaginary part $\operatorname{Im}\Pi(s=E^{2})$ can be calculated in the
framework of a given QFT. In the easiest case, it is evaluated at tree-level,
see e.g. Ref. \cite{lupo} and Section 3.3 for a concrete example.
Higher orders can also be evaluated but are technically more involved, even
though in general the correction with respect to the tree-level result is small
\cite{schneitzer}.

The time evolution of the state in QFT reads \cite{duecan,nonexpqft}
\begin{equation}
a_{S}(t)=\frac{i}{2\pi}\int_{-\infty}^{+\infty}G_{S}(s)e^{-i\sqrt{s}t}%
ds=\int_{s_{th}}^{\infty}d_{S}(s)e^{-i\sqrt{s}t}ds \text{ .}
\end{equation}
Note, the first integral extends from $-\infty$ to $+\infty,$ i.e. it includes
also unphysical values of the variable $s.$

\subsection{Standard relativistic approximation}

In order to obtain possible expressions for the relativistic BW distribution, let us consider%
\begin{equation}
\operatorname{Im}\Pi(s)=M\Gamma\theta(s)\text{ ,}%
\end{equation}
which closely resembles the BW case of Sec. 2 by including a lowest threshold
at zero. Of course, $\operatorname{Im}\Pi(s)=M\Gamma\theta(s)$ is itself not
physical, but - as we shall see - additional approximations are needed to
obtain the standard relativistic BW distribution.

The loop function takes the form  (detailed studies of the structure of the loops have been carried out before on a case by case basis, see, {\it e.g,}  \cite{Hoferichter:2009gn}):
\begin{equation}
\Pi(s)=-\int_{0}^{\Lambda^{2}}\frac{\mathrm{ds}^{\prime}}{\pi}\frac{M\Gamma
}{s-s^{\prime}+i\varepsilon}=\frac{M\Gamma}{\pi}\left[  \ln(s^{\prime
}-s)\right]  _{0}^{\Lambda^{2}}=\frac{M\Gamma}{\pi}\ln\frac{s-\Lambda^{2}}%
{s}=\frac{M\Gamma}{\pi}\ln\frac{-\Lambda^{2}}{s} \text{ .}
\end{equation}
By including a subtraction constant $C=-\frac{M\Gamma}{\pi}\ln
\frac{M^{2}}{\Lambda^{2}}$ and by taking 
$\Lambda\rightarrow\infty$
we get: 
\begin{equation}
\Pi(s)=\frac{M\Gamma}{\pi}\ln\frac{-\Lambda^{2}}{s}-C=\frac{M\Gamma}{\pi}%
\ln\frac{-\Lambda^{2}}{s}+\frac{M\Gamma}{\pi}\ln\frac{M^{2}}{\Lambda^{2}%
}=\frac{M\Gamma}{\pi}\ln\frac{-M^{2}}{s} \text { ,} 
\end{equation}
which assures that $\operatorname{Re}\Pi(s=M^{2})=0,$ thus the mass is still
$M$ even after including the one-particle self-energy. The important point is that the real part of the loop $\operatorname{Re}\Pi(s)=\frac{M\Gamma}{\pi}\ln\frac{M^{2}}{s}$ is not a simple constant. 

The propagator reads
\begin{equation}
G_{S}(s)=\frac{1}{s-M^{2}+\frac{M\Gamma}{\pi}\ln\frac{M^{2}}{s}+iM\Gamma
\theta(s)+i\varepsilon}\text{ .}%
\end{equation}
Note, the pole has no simple algebraic solution.
The corresponding spectral function takes the form
\begin{equation}
d_{S}(s)=-\frac{1}{\pi}\operatorname{Im}[G_{S}(s)]=\frac{1}{\pi}\frac{M\Gamma
}{(s-M^{2}+\frac{M\Gamma}{\pi}\ln\frac{M^{2}}{s})^{2}+(M\Gamma)^{2}}\theta(s)
\text{ ,}
\end{equation}
which is correctly normalized to 1, as it should. 

The propagator for the relativistic BW approximation is obtained by
approximating the previous propagator upon setting the real part of the loop
artificially to zero, thus finding:
\begin{equation}
G_{S}(s)=G^{BW}_{S}(s)=\frac{1}{s-M^{2}+iM\Gamma+i\varepsilon}\text{ ,} \label{bwprop}%
\end{equation}
whose corresponding pole is%
\begin{align}
s_{pole}  &  =M^{2}-iM\Gamma\rightarrow\\
\sqrt{s_{pole}}  &  =\sqrt{M}\sqrt{M-i\Gamma}=\sqrt{\frac{M}{2}}\sqrt
{M+\sqrt{M^{2}+\Gamma^{2}}}-i\sqrt{\frac{M}{2}}\sqrt{-M+\sqrt{M^{2}+\Gamma
^{2}}}
\end{align}
Indeed, the form of Eq. (\ref{bwprop}) is directly used in many applications, e.g. Refs. \cite{sato,peters,qi,back} and refs. therein.
The rBW spectral function as function of $s$ reads%
\begin{equation}
d_{S}^{\text{rBW}}(s)=\frac{1}{\pi}\frac{M\Gamma}{(s-M^{2})^{2}+(M\Gamma)^{2}%
}\theta(s) \text{ ,}\label{rBWM}
\end{equation}
to be understood for $s\geq0$, hence the step function has been added. 
The related spectral function, as a function of the energy $E$ takes the form %
\begin{equation}
d_{S}(E)=d_{S}^{\text{rBW}}(E)=\frac{2E}{\pi}\frac{M\Gamma}{(E^{2}-M^{2}%
)^{2}+(M\Gamma)^{2}}\theta(E)\text{ ,} \label{bw1}
\end{equation}
that contains an unphysical jump at $E=0$. We refer e.g. Refs. \cite{alicerbw,starrbw,vovchenko,pythia} for a practical use of this function\footnote{In some cases the step function is not explicitly introduced, but is always implicitly present, since in a relativistic framework $E<0$ is meaningless.}  This distribution function is in fact an approximation where, the factor of $E\Gamma$ in the numerator has been replaced\footnote{For the analysis of the resonances without this approximation, see Appendix \ref{rBWE}.} by $M\Gamma$.

The normalization to 1 is lost due to the employed approximation: the real part is not a constant and therefore cannot be completely subtracted.
Of course, one may normalize the distribution of Eq. (\ref{bw1}) by \textquotedblleft brute force \textquotedblright upon considering%
\begin{equation}
d_{S}^{\text{rBW-norm}}(E)=\frac{1}{\frac{1}{2}+\frac{1}{\pi}\arctan
\frac{M}{\Gamma}}\frac{2E}{\pi}\frac{M\Gamma}{(E^{2}-M^{2})^{2}+(M\Gamma)^{2}%
}\theta(E)\text{ ,}%
\end{equation}
but this step is not mathematically consistent. 
wit
Curiously, as we shall show in Sec. 5, the rBW fits to the data in a similar way as the BW. As shown in the numerical examples, the two plots are not discernible to the naked eye. However, the pole positions and the decay widths are marginally different.

\subsection{A scalar QFT}

In this context it is worth reminding the results for the QFT case in which a
scalar particle decays into two (pseudo)scalars. The corresponding interacting
part of the Lagrangian takes the form \cite{lupo}
\begin{equation}
\mathcal{L}_{int}=gS\varphi^{2}\text{,}%
\label{sphi2}
\end{equation}
out of which (at one-loop):
\begin{equation}
\operatorname{Im}\Pi(s)=g^{2}\frac{\sqrt{\frac{s}{4}-m^{2}}}{4\pi\sqrt{s}%
}\theta(s-4m^{2})\text{ .}\label{imqft}%
\end{equation}
The on-shell decay width is given by
\begin{equation}
\Gamma=g^{2}\frac{\sqrt{\frac{M^{2}}{4}-m^{2}}}{4\pi M^{2}}\text{ .}%
\end{equation}
The loop takes the form%
\begin{align}
\Pi(s) &  =-g^{2}\int_{4m^2}^{\Lambda^{2}}\frac{\mathrm{ds}^{\prime}}{\pi}%
\frac{\frac{\sqrt{\frac{s^{\prime}}{4}-m^{2}}}{4\pi\sqrt{s^{\prime}}}%
}{s-s^{\prime}+i\varepsilon}\\
&  =g^{2}\frac{-\sqrt{4m^{2}-s}}{4\pi^{2}\sqrt{s}}\arctan\left(  \frac
{\Lambda\sqrt{s}}{\sqrt{\Lambda^{2}+m^{2}}\sqrt{4m^{2}-s}}\right)
-\frac{g^{2}}{4\pi^{2}}\log\left(  \frac{m}{\Lambda+\sqrt{\Lambda^{2}+m^{2}}%
}\right)  \text{.}%
\end{align}
Let us then take the limit $\Lambda\rightarrow\infty$ and introduce a
subtraction constant in such a way that $\operatorname{Re}[\Pi(M^{2})]=0$. Upon employing $\arctan\left(  z\right)  =\frac{1}{2i}%
\log\left(  \frac{i-iz}{i+iz}\right)  $) one has:
\begin{equation}
\Pi(s)=g^{2}\frac{\sqrt{\frac{s}{4}-m^{2}}}{4\pi^{2}\sqrt{s}}\log\left(
\frac{\sqrt{s-4m^{2}}-\sqrt{s}}{\sqrt{s-4m^{2}}+\sqrt{s}}\frac{\sqrt
{M^{2}-4m^{2}}+M}{M-\sqrt{M^{2}-4m^{2}}}\right) \text{ .}
\end{equation}
Then, the propagator is 
perfectly defined and the spectral function is correctly normalized. On the other
hand, the real part of the loop is a nontrivial function with a cusp at
$4m^{2},$ which is nonzero above threshold \cite{lupo}.

The present model has a great advantage with respect to the BW and rBW cases, since it
correctly implements the lowest energy threshold and it arises from a - albeit
simple - QFT approach. Yet, it would be awkward to use such a function in 
fitting experimental data for various reasons:

1) the QFT model of Eq. (\ref{sphi2}) can be only considered as a first approximation for physical resonances.

2) Even in the simplest decays of a scalar particles into two (pseudo)scalar ones there can be
additional terms, such as $S\partial_{\mu}\varphi\partial^{\mu}\varphi$
\cite{boglione,achasov,lupoder,wolkak,guo}. For instance, this is the case for the spectral function of the
resonances $a_{0}(980),$ $a_{0}(1450),$ $K_{0}^{\ast}(700),$ and $K_{0}^{\ast
}(700)$.

3) Decaying particles (especially hadrons) involve also particle with
spins.\ For the use of spectral functions for various vector particles,  see for instance Ref. \cite{psi3770} for the
vector field $\psi(3770),$ Ref. \cite{x3872,kalashnikova,zhoux3872,Wang:2020tpt,Artoisenet:2010va,Braaten:2007dw} for the
$X(3872),$ Ref. \cite{piotrowska} for the state $\psi(4040),$ Ref.
\cite{y4260} for $Y(4260),$ and Ref. \cite{Cao:2019wwt} for the $\psi(4660).$ 

4) Typically, when derivative and/or higher spin are considered, the divergence of the loop integral is not just logarithmic, hence
additional subtractions or cutoff functions are needed (note, one may also use
explicit form factors, which however corresponds to nonlocal Lagrangians, see
e.g. Ref. \cite{terning,nonlocal} and refs. therein).

5) Quite interestingly, the form of Eq. (\ref{imqft}) has been used in a
variety of applications, see for instance the recent paper by the ALICE
collaboration \cite{alicerho0} (but the coprresponding real part has been neglected).

\section{A useful relativistic Flatt\'{e}-like spectral function: the Sill}

In this section we present a relativistic spectral function that properly includes the threshold (with no unphysical jump) and is always correctly normalized for both equal or unequal masses in the final state. Moreover, it is easily extendable to the multichannel case and to decay chains. 

\subsection{Definition and properties}

Let us consider a certain decay whose threshold is $s_{th}=E_{th}^{2}.$ For a
two-body decay, one has $E_{th}=m_{1}+m_{2}=\sqrt{s_{th}}.$ We then \textit{assume} the
following simple imaginary part of the loop function%

\begin{equation}
\operatorname{Im}\Pi(s)=\sqrt{s-s_{th}}\tilde{\Gamma}\theta(s-s_th) \label{assumption}%
\text{ ,}
\end{equation}
where $\tilde{\Gamma}$ is a rescaled width linked to the standard decay width
$\Gamma$ via the relation $\Gamma M=\tilde{\Gamma}\sqrt{M^{2}%
-E_{th}^{2}},$ see below. Notice that in this context the function $\Gamma(s)$
is actually not a constant. In fact, from%
\begin{equation}
\operatorname{Im}\Pi(s)=\sqrt{s}\Gamma(s)
\end{equation}
it follows that (for $s\geq s_th$)
\begin{equation}
\Gamma(s)=\frac{\sqrt{s-s_{th}}}{\sqrt{s}}\tilde{\Gamma}\text{ ,} \label{gammaofs}%
\end{equation}
which on-shell reduces to
\begin{equation}
\Gamma=\Gamma(s=M^{2})=\tilde{\Gamma}\frac{\sqrt{M^{2}-E_{th}^{2}}}{M}\text{
.}\label{gammatilde}
\end{equation}
Yet, even if not a constant, the width $\Gamma(s)$ saturates for large $s$ to a constant equal to $\tilde{\Gamma}.$ For $M$ sufficiently larger than $E_{th},$
it follows that $\Gamma\simeq\tilde{\Gamma}.$

\begin{table}[t]
    \centering
    \begin{tabular}{|c|c|c|c|c|c|c|c|}
    \hline
     Distribution & $g_\rho$ (MeV) & $b_\rho$ & $g_{a_1}$ (MeV) & $b_{a_1}$ & $g_{K^\ast}$ (MeV) & $b_{K^\ast}$ & $g_\Delta$ (MeV)\\\hline
     Nonrelativistic BW & $622.9$ & $0.012$ & $473.7$ & $0.233$ & $39.1\times 10^6$ & $107742.2$ & $139.2$ \\
     Relativistic BW & $622.3$ & $0.019$ & $479.5$ & $0.234$ & $73.5\times 10^6$ & $109122.0$ & $139.3$\\
     Sill & $597.8$ & $0.038$ & $757.9$ & $ 0.037$ & $39.0\times 10^6$ & $108217.0$ & $147.7$\\\hline
    \end{tabular}
    \caption{Parameters used to calculate the spectral function as given in eq.(\ref{fiteq}) of the text.}
    \label{fittab}
\end{table}

The loop takes the form
\begin{align}
\Pi(s)  &  =-\int_{s_{th}}^{\Lambda^{2}}\frac{\mathrm{ds}^{\prime}}{\pi}%
\frac{\sqrt{s-s_{th}}\tilde{\Gamma}}{s-s^{\prime}+i\varepsilon}\\
&  =\frac{2}{\pi}\tilde{\Gamma}\left(  \sqrt{\Lambda^{2}-s_{th}}+\frac
{\sqrt{s-s_{th}}}{2}\ln\frac{\sqrt{s-s_{th}}-\sqrt{\Lambda^{2}-s_{th}}}%
{\sqrt{s-s_{th}}+\sqrt{\Lambda^{2}-s_{th}}}\right)
\end{align}
By taking the limit $\Lambda\rightarrow\infty$ together with one subtraction,
the loop function reduces to
\begin{equation}
\Pi(s)=i\tilde{\Gamma}\sqrt{s-s_{th}}%
\end{equation}
which is then only imaginary for $s \geq s_{th}$ (conversely, it is only real below threshold).

Note, the function $\Pi(s)$ for complex values of the
variable $s$ is a multivalued function with two Riemann sheets (RS);
the branch cut is taken for $s\geq s_{th}\geq0.$ The expression above
is therefore on the I RS, while the expression on the II RS is simply given by
$\Pi_{II}(s)=-\Pi(s)$. For real $s<s_{th}$ the function
$\Pi(s)$ is real, thus the Schwarz reflection principle is fulfilled.
The spectral representation of $\Pi(s)$ (obtained with one
subtraction) takes the form: %
\begin{equation}
\Pi(s)=-\frac{s-s_{th}}{\pi}\int_{s_{th}}^{\infty}ds^{\prime}\frac{1}%
{\sqrt{s^{\prime}-s_{th}}}\frac{1}{s-s^{\prime}+i\varepsilon}\text{ }%
\end{equation}
valid for any $s$ besides the cut $(s_{th},\infty).$

The propagator takes the form
\begin{equation}
G_{S}(s)=\frac{1}{s-M^{2}+i\tilde{\Gamma}\sqrt{s-s_{th}}+i\varepsilon}\text{,
}%
\end{equation}
whose relevant pole (on the II RS)is%
\begin{equation}
s_{pole}=M^{2}-\frac{\tilde{\Gamma}^{2}}{2}-i\sqrt{(M^{2}-s_{th})\tilde
{\Gamma}^{2}+\frac{\tilde{\Gamma}^{4}}{4}} \text{ .} %
\end{equation}
Note, for $\tilde{\Gamma}^{2}$ sufficiently smaller than $M^{2}-s_{th},$ the
pole of $s$ can be approximated as
\begin{equation}
s_{pole}\simeq M^{2}-i\sqrt{(M^{2}-s_{th})}\tilde{\Gamma}=M^{2}-iM\Gamma\text{ ,}%
\end{equation}
just as in the rBW case. 

Next, the spectral function reads%
\begin{equation}
d_{S}(s)=-\frac{1}{\pi}\operatorname{Im}\frac{1}{s-M^{2}+i\tilde{\Gamma}%
\sqrt{s-s_{th}}+i\varepsilon}=\frac{1}{\pi}\frac{\sqrt{s-s_{th}}\tilde{\Gamma
}}{(s-M^{2})^{2}+(\sqrt{s-s_{th}}\tilde{\Gamma})^{2}}\theta(s-s_{th}) \text{ .}
\end{equation}
or, as function of the energy $E$, as:%

\begin{equation}
d_{S}(E)=d_{S}^{\text{ Sill}}(E)=\frac{2E}{\pi}\frac{\sqrt{E^{2}-E_{th}^{2}%
}\tilde{\Gamma}}{(E^{2}-M^{2})^{2}+\left(  \sqrt{E^{2}-E_{th}^{2}}%
\tilde{\Gamma}\right)  ^{2}}\theta(E-E_{th}) \text{ .}
\end{equation}

The normalization 
\begin{equation}
\int_{E_{th}}^{+\infty }\mathrm{dE}~\!d_{S}^{\text{Sill}}(E)=1
\end{equation}%
for any $E_{th},$ $M$, and $\tilde{\Gamma}$ is a consequence of the proper
treatment of the real part of the loop according to the general proof
reported in Refs. \cite{leerev,lupofermions}. Of course, it can be also directly
proven for the specific spectral function $d_{S}^{\text{Sill}}(E).$

\begin{figure}[t]
    \centering
    \includegraphics[scale=0.4]{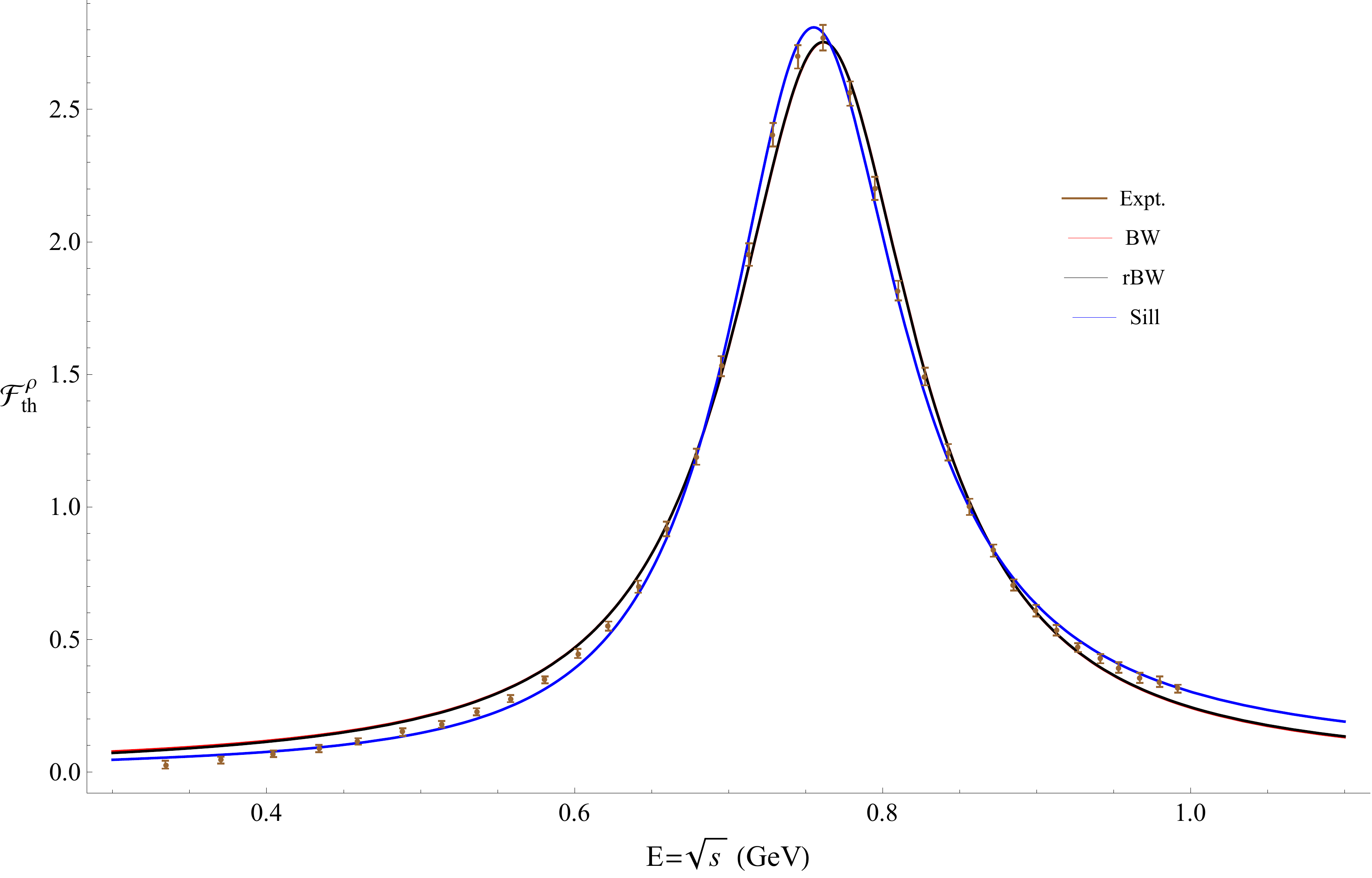}
    \caption{The spectral function for the $\rho(770)$. Experimental data corresponds to the vector channel of the $\tau\to\pi\pi\nu_\tau$ decay reported in \cite{aleph}.}
    \label{rhoplot}
\end{figure}

\begin{table}[h]
    \centering
    \begin{tabular}{|c|c|c|c|c|}
        \hline
         Distribution & $M$ (MeV) & $\Gamma$ (MeV) & $\chi^2/\rm{d.o.f}$ & $\sqrt{s_{pole}} (\text{MeV})$\\\hline
         Nonrelativistic BW & $761.64 \pm 0.32$ & $144.6 \pm 1.3$ & $10.16$ & $761.6 -i\, 72.3$\\
         Relativistic BW & $758.1 \pm 0.33$ & $145.2\pm 1.3$ & $9.42$ &$761.5 -i\, 72.3$\\
         Sill & $755.82\pm 0.33$ & $137.3\pm 1.1$ & $3.52$ & $751.7 -i\, 68.6 $ \\\hline
    \end{tabular}
    \caption{Mass and widths of $\rho(770)$ fitted using the three distributions discussed in the text, their error estimates, and the poles.}
    \label{rhotab}
\end{table}

Some comments are in order:

1) The Sill distribution correctly reduces to the rBW, $d_{S}^{\text{sill}}(E)\rightarrow d_{S}^{\text{rBW}}(E)$,   upon substituting  
\begin{equation}
\sqrt{E^{2}-E_{th}^{2}}\tilde{\Gamma}\simeq \sqrt{M^{2}-E_{th}^{2}}\tilde{%
\Gamma}=M\Gamma 
\end{equation}%
both in the numerator and denominator. This is allowed when the
width is small. 
It is also interesting to notice that $d_{S}^{\text{Sill}}(E)$ scales for large $E$ as $E^{-2}$, just as in the BW case (and not as $E^{-3}$ as in the rBW of Eq. (\ref{rBW})). 

2) A comparison with the QFT case of Sec. 3.2 is useful. In that case, one
had \begin{equation}
\operatorname{Im}\Pi(s)=g^{2}\frac{\sqrt{\frac{E^{2}}{4}-m^{2}}}{4\pi E}%
=g^{2}\frac{\sqrt{E^{2}-E_{th}^{2}}}{8\pi E}\text{ ,}%
\end{equation}
hence the difference is
clear: the additional energy dependence in the denominator does not
allow for a constant real part. Yet, the numerator is the same of the Sill distribution in the case of equal masses. The QFT case of Section 3.3
 translates into the Sill form upon the substitution
\begin{equation}
E\rightarrow M\text{ in the denominator,}
\end{equation}
a step which is necessary to get a vanishing real part of the self-energy. 
Here, it must be stressed that the modification of the high-energy tail does not constitute a drawback for the Sill\footnote{Actually, the  imaginary part of the loop should vanish for large energies, a fact which is not described by either the simple QFT in Sec. 3.2 or the Sill.}, since the high-energy region has a small influence on the behavior of the distribution close to the peak.

On the other hand, the fact that for the Sill $\Gamma(s\rightarrow\infty)=\tilde{\Gamma}$ (i.e. it is a constant for high energies) implies that in this respect the Sill closely resembles (r)BW. Also, for the Sill the real part of the self-energy vanishes, in agreement with BW (but not with rBW). Colloquially speaking, we might say that the Sill is \textquotedblleft closer\textquotedblright \ 
to the BW than rBW itself, and it achieves that by properly taking into account the threshold.

3) Next, let us consider the case of the QFT decays in two distinct particles with different mass.
The corresponding scalar QFT Lagrangianes is $\mathcal{L}_{int}=gS\varphi
_{1}\varphi_{2},$ then the imaginary part reads%
\begin{equation}
\operatorname{Im}\Pi(s)=g^{2}\frac{k(E)}{8\pi E}\text{ ,}%
\end{equation}
with
\begin{equation}
k(E)=\frac{1}{2}\sqrt{\frac{E^{4}+(m_{1}^{2}-m_{2}^{2})^{2}-2(m_{1}^{2}%
+m_{2}^{2})E^{2}}{E^{2}}}.
\end{equation}
Thus, the Sill distribution proposed in Eq. (\ref{Sill}) amounts to:
\begin{align}
k(E)  &  \rightarrow\sqrt{E^{2}-E_{th}^{2}}\text{ in the numerator with $E_{th}=m_1+m_2$;}\\
E  &  \rightarrow M\text{ in the denominator.}%
\end{align}
Indeed, the behaviors of the functions $k(E)$ and $\sqrt{E^{2}-E_{th}^{2}}$ are
quite similar for $E>E_{th},$ but the latter is easier to implement. Moreover, in the case of unequal masses, the choice of $\operatorname{Im}\Pi(s)$ as proportional to $k(E)$ does {\it not} lead to a properly normalized spectral function (even when substituting the factor $E$ in the denominator with a constant $M$). Since this requirement is central to our approach, the choice implemented in the Sill is both simpler and theoretically favoured. 

4) Finally, it should be noticed that in some cases also the generalized form
involving the angular momentum $l$ is introduced as
\begin{equation}
\operatorname{Im}\Pi(s)=g^{2}\frac{k(E)^{2l+1}}{4\pi E}\text{ ,}%
\end{equation}
yet for $l>0$ the divergence of the loop function is too severe and
the normalization of the spectral function is lost. This aspect shows that
some sort of cutoff function is needed for a proper treatment of the problem
and for a meaningful probabilistic interpretation of the mass/energy
distribution. This is why we do not attempt to include $l$ into the proposed
phenomenological description introduced in Eq. (\ref{assumption}) and to the
spectral function $d_{S}^{\text{Sill}}(E)$ of Eq. (\ref{Sill}). As we shall see later on in concrete examples, the Sill fairs well also when higher values of the angular momentum $l$ are actually involved. 

\subsection{Extension to the multi-channel case\label{multiSill}}

In many cases unstable resonances decay in more than a single decay
channel \cite{pdg}.

We consider as an illustrative example the case of an unstable state $S$ with
two open decay channels with rescaled partial decay widths $\tilde{\Gamma}_{1}$ and
$\tilde{\Gamma}_{2}$ (linked to the partial decay width via $\Gamma_{k}%
=\tilde{\Gamma}_{k}\sqrt{M^{2}-E_{k,th}^{2}}/M)$ as well as the two thresholds
$s_{1,th}$ and $s_{2,th}\geq s_{1,th}$. 
The propagator is: 

\begin{equation}
G_{S}(s)=\frac{1}{s-M^{2}+i\tilde{\Gamma}_{1}\sqrt{s-s_{1,th}}+i\tilde{\Gamma
}_{2}\sqrt{s-s_{2,th}}+i\varepsilon}\text{, }%
\end{equation}
Care is needed when deriving the spectral function, since different regions emerge: 
\begin{equation}
d_{S}(s)=-\frac{1}{\pi}\operatorname{Im}[G_{S}(s)]=\left\{
\begin{array}
[c]{c}%
\frac{1}{\pi}\frac{\tilde{\Gamma}_{1}\sqrt{s-s_{1,th}}+\tilde{\Gamma}_{2}%
\sqrt{s-s_{2,th}}}{\left(  s-M^{2}\right)  ^{2}+(\tilde{\Gamma}_{1}%
\sqrt{s-s_{1,th}}+\tilde{\Gamma}_{2}\sqrt{s-s_{2,th}})^{2}}\text{ for
}s>s_{2,th}\\
\frac{1}{\pi}\frac{\tilde{\Gamma}_{1}\sqrt{s-s_{1,th}}}{\left(  s-M^{2}%
-\tilde{\Gamma}_{2}\sqrt{s-s_{2,th}}\right)  ^{2}+(\tilde{\Gamma}_{1}%
\sqrt{s-s_{1,th}})^{2}}\text{ for }s_{1,th}\leq s\leq s_{2,th}\\
0\text{ for }s<s_{1,th}%
\end{array}
\right.  .
\end{equation}
Clearly, the normalization still holds from the lowest threshold:
$\int_{s_{1,th}}^{\infty}d_{S}(s)ds=1.$
This is a consequence of unitarity, which is a necessary feature of the unstable
state S. Thus, the present generalization of the Sill is a simple,
straightforward extension to the case of multiple decay channels that is
both consistent and sufficient for our purposes.
However, we stress that the present multichannel extension refers only to the
shape of the spectral function of the unstable state in the framework of the
used approximation(s), but not to the whole multichannel scattering problem
that involve scattering amplitudes, see for instance \cite{Braaten:2007dw}.


The extension to the $N$ channels is straightforward:
\begin{equation}
G_{S}(s)=\frac{1}{s-M^{2}+i\sum_{k=1}^{N}\tilde{\Gamma}_{k}\sqrt{s-s_{k,th}%
}+i\varepsilon}%
\end{equation}
with%
\begin{equation}
\tilde{\Gamma}_{k}=\Gamma_{k}\frac{M}{\sqrt{M^{2}-E_{k,th}^{2}}}\text{ and
}s_{1,th}=E_{1,th}^{2}\leq s_{2,th} \leq ...\leq_{N,th}=E_{N,th}^{2}\text{ .}%
\end{equation}
The spectral function is
\begin{equation}
d_{S}(s)=-\frac{1}{\pi}\operatorname{Im}[G_{S}(s)] \text{ ,}  
\end{equation}
which takes the explicit form for e.g. $s_{Q,th}\leq s \leq s_{Q+1,th}$:
\begin{equation}
d_{S}(s)=\frac{1}{\pi}\frac{\sum_{k=1}^{Q}\tilde{\Gamma}_{k}\sqrt{s-s_{k,th}}%
}{\left(  s-M^{2}-\sum_{k=Q+1}^{N}\tilde{\Gamma}_{k}\sqrt{s-s_{k,th}}\right)
^{2}+\left(  \sum_{k=1}^{Q}\tilde{\Gamma}_{k}\sqrt{s-s_{k,th}}\right)^{2}%
}
\end{equation}
with $Q=1,...,N$. The normalization is fulfilled also in this general case.
Note, in other intervals of $s$ one should carefully check which pieces appear in the numerator and which ones affect the mass. 

Within this context it is also useful to introduce the partial $i$-th spectral
function as
\begin{equation}
d_{S}^{(i)}(s)=\frac{1}{\pi}\Gamma_{i}\sqrt{s-s_{i,th}}\theta(s-s_{i,th}) |G_{S}(s)|^2 \text{ .}  
\end{equation}
The latter reads explicitly (for $s_{Q,th}\leq s \leq s_{Q+1,th}$):

\begin{equation}
d_{S}^{(i)}(s)=\frac{1}{\pi}\frac{\tilde{\Gamma}_{i}\sqrt{s-s_{i,th}}}{\left(
s-M^{2}-\sum_{k=Q+1}^{N}\tilde{\Gamma}_{k}\sqrt{s-s_{k,th}}\right)
^{2}+\left(  \sum_{k=1}^{Q}\tilde{\Gamma}_{k}\sqrt{s-s_{k,th}}\right)^{2}
} \text{,}%
\end{equation}
where only the $i$-th channel is present in the numerator (the denominator
being unchanged). The quantity $d_{S}^{(i)}(s)ds$ may be interpreted as the
probability that the unstable state has a squared energy between $s$ and $s+ds$
and decays subsequently in the $i$-th channel. 
As a consequence the integral
\begin{equation}
\int_{s_{i,th}}^{\infty}d_{S}^{(i)}(s)\mathrm{ds}%
\end{equation}
can be seen as the branching ratio for the decay in the $i$-th channel. 
Of course, the sum of all branching ratios is one. 

In Sec. 5 we shall apply the multichannel Sill to the resonances $f_2(1270)$ and $a_0(980)$.

\subsection{Convolution of the Sill}

Quite often, decay chains take place. As an example, let us consider the unstable state $S$
decaying into
\begin{equation}
S\rightarrow\varphi_{1}\varphi_{2}\text{ ,}%
\end{equation}
where $\varphi_{1}$ is stable but the field $\varphi_{2}$ is 
itself unstable and decays into
\begin{equation}
\varphi_{2}\rightarrow\varphi_{3}\varphi_{4} \text{ .}%
\end{equation}
The spectral function of the state $\varphi_{2}$ reads (using the Sill for consistency, for other approaches see Refs. \cite{Nauenberg:1962aa,Hanhart:2010wh,Zhang:2021hcl}) %
\begin{equation}
d_{\varphi_{2}}(s)=-\frac{1}{\pi}\operatorname{Im}\frac{1}{s-m_{2}^{2}%
+i\tilde{\Gamma}_{2}\sqrt{s-(m_{3}+m_{4})^{2}}+i\varepsilon} \text{ ,}%
\end{equation}
then the convoluted spectral function of the state $S$ takes this aspect into
account.
Next, we consider
\begin{equation}
\tilde{d}_{S}(s,s^{\prime})=-\frac{1}{\pi}\operatorname{Im}\frac{1}%
{s-M^{2}+i\tilde{\Gamma}\sqrt{s-(m_{1}+\sqrt{s^{\prime}})^{2}}+i\varepsilon} \text{ ,}%
\end{equation}
in which the mass of the field $\varphi_{2}$ is taken as the running quantity $\sqrt{s^{\prime}}$.
Then the convoluted Sill reads
\begin{equation}
d_{S}^{\text{conv-sill}}(s)=\int_{(m_{3}+m_{4})^{2}}^{\infty}ds^{\prime}%
d_{\varphi_{2}}(s^{\prime})\tilde{d}_{S}(s,s^{\prime}) \text{ .}
\end{equation}
Intuitively, we are taking the average of all possible masses of the unstable
field $\varphi_{2}$ weighted by $d_{\varphi_{2}}(s^{\prime})$. Note, for
$\tilde{\Gamma}_{2}\rightarrow0$ one has $d_{\varphi_{2}}(s)=\delta
(s-m_{2}^{2}),$ thus one reobtains the expected spectral function. The
normalization of the convoluted Sill is also straightforward:
\begin{align}
\int_{(m_{1}+m_{2}+m_{3})^{2}}^{\infty}dsd_{S}^{\text{conv-sill}}(s) &
=\int_{(m_{1}+m_{2}+m_{3})^{2}}^{\infty}ds\int_{(m_{3}+m_{4})^{2}}^{\infty
}ds^{\prime}d_{\varphi_{2}}(s^{\prime})\tilde{d}_{S}(s,s^{\prime})\\
&  =\int_{(m_{3}+m_{4})^{2}}^{\infty}ds^{\prime}d_{\varphi_{2}}(s^{\prime
})\int_{(m_{1}+m_{2}+m_{3})^{2}}^{\infty}ds\tilde{d}_{S}(s,s^{\prime})\\
&  =\int_{(m_{3}+m_{4})^{2}}^{\infty}ds^{\prime}d_{\varphi_{2}}(s^{\prime})=1 \text{ ,}
\end{align}
where we used that%
\begin{equation}
\int_{(m_{1}+m_{2}+m_{3})^{2}}^{\infty}ds\tilde{d}_{S}(s,s^{\prime}%
)=\int_{(m_{1}+\sqrt{s^{\prime}})^{2}}^{\infty}ds\tilde{d}_{S}(s,s^{\prime})=1 \text{ .}
\end{equation}

Of course, one can also generalize the study of decay chains to the case of multiple decays. For specific examples, see the discussion about the resonances $a_1(1260)$ and $f_2(1270)$ in Sec. 5.

In general, the convoluted Sill distribution is more involved and requires
more efforts to be used in a fit than the plain Sill and it can be usually
omitted in first approximation. Nevertheless, in presence of precise data it can be an additional tool in some selected 
cases since it represents a straightforward, self-contained, and normalized
extension of the Sill distribution.

\section{Numerical examples}

The spectral functions studied in the previous sections (BW, rBW, and Sill) enter the relation
\begin{equation}
    \mathcal{F}_{th} = g~ d_S(E) + b \text{ ,} 
    \label{fiteq}
\end{equation}
where $\mathcal{F}_{th}$ represents the theoretical quantity that can be compared to data, $g$ is a scaling factor (typically a coupling constant, but that depends on the considered experiment), and $b$ is the background. Since $g$ is often an important physical quantity, it is evident that the correct normalization of the spectral function $d_S(E)$ is relevant.

The values of these coefficients used in the present work are listed in Table. \ref{fittab}. Indeed, for the $\rho$ and the $a_1$ cases, the coupling constants can be seen as physical couplings, as explained in Ref. \cite{aleph}.  On the other hand, for the $K^*$, the experimental value refer to row counts. 
The parameters listed in the subsequent tables were estimated by minimizing the $\chi^2$  defined as

\begin{equation}
    \chi^2 = \sum \left(\frac{\mathcal{F}_{th}-\mathcal{F}_{expt}}{\Delta\mathcal{F}_{expt}}\right)^2 \text{.}
\end{equation}
for the respective distributions.
The errors in the parameters were calculated using the Hesse matrix method as explained in Appendix \ref{hesse}. 

We stress that the examples that we shall present are illustrative and aim to show the utility of the Sill distribution as a useful function to fit data in first approximation. Yet, for each of the presented resonances, more detailed and advanced studies have been performed ({\it e.g,} see \cite{Dudek:2016cru,Briceno:2017qmb} for a review of various methods as well as the PDG review on {\it Resonances} \cite{pdg}). The Sill is not a substitute of such advanced studies, but a simple useful tool to study resonances by taking into account threshold effects.

\subsection{The $\rho$ meson}\label{rhodis}

The three functions $\mathcal{F}$ for the $\rho(770)$, fitted using the three distribution functions (BW, rBW, Sill), along with the experimental data \cite{aleph} are given in Fig. \ref{rhoplot}. The mass and the width used to fit the data, their error estimates, and the pole position are listed in Table \ref{rhotab}. 

We see that the BW and rBW distributions fit the data poorly (as indicated by the higher value of the reduced $\chi^2$), compared to the Sill. 
There is also a marked difference between the mass and the width of the $\rho(770)$ estimated by the Sill distribution compared to the other distributions. The difference between the three distributions is not so large close to the peak because the $2\pi$ threshold is far away. Yet, as expected, the Sill fits the left tail clearly better.  This behavior plays an important role in the case of resonances with the poles closer to the threshold (e.g. the $a_1(1260)$ discussed next).

We notice that the Sill can describe the $\rho$ spectral function relatively well even though the correct angular momentum is not properly taken into account. Namely, for the $\rho$, one has actually the scaling $k(E)^{3}$. 
As the detailed study of Ref. \cite{Gounaris:1968mw} shows, a proper implementation of the angular momentum is important for an accurate description of data. Yet, the effects of the threshold are -in the first approximation- more relevant than the angular momentum ones.

The study in Ref. \cite{Gounaris:1968mw} shows another important point. In that work, the real part of the loop was evaluated by including two subtractions to guarantee convergence of the one-loop function. The treatment is more involved and definitely preferable whenever accurate data are available. On the other hand, the Sill distribution discussed here is much simpler since the real part can be (consistently) set to zero.

In conclusions, the Sill does not represent a substitute to a detailed study of a given resonance, but represents a simple function to start with, especially if some thresholds are close. It seems therefore that the role of the left energy threshold is enough for a fair description of data. These are good news, since the neglect (in first approximation) of the angular momentum renders the treatment of line shapes easier and systematically possible within the Sill function presented here.


\bigskip

\subsection{The $a_{1}(1260)$ meson}
\begin{table}[h]
    \centering
    \begin{tabular}{|c|c|c|c|c|}
        \hline
        Distribution & $M$ (MeV) & $\Gamma$ (MeV) & $\chi^2/\rm{d.o.f}$ & $\sqrt{s_{pole}} (\text{MeV})$\\\hline
         Nonrelativistic BW & $1165.6\pm1.2$ & $415\pm 15$ & $4.31$& $1166 -i\, 208$\\
         Relativistic BW & $1146.5\pm 1.6$ & $424 \pm 16$ & $4.25$& $1165 -i\, 209 $\\
         Sill & $1181.3 \pm 3.4$ & $539 \pm 27$ & $3.52$ & $1046 -i\, 250 $\\\hline
    \end{tabular}
    \caption{Mass and width of $a_1(1260)$ fitted using the three distributions discussed in the text, their error estimates, and the poles (as described in the text).}
    \label{a1tab}
\end{table}
As discussed before, the non-relativistic BW distribution is ill-equipped to handle resonances lying close to the threshold. On the other hand, even though the relativistic BW distribution can be modified to include a threshold, the latter has to be artificially imposed. This limitation of the BW distributions appear stark in case of resonances like the $a_1(1260)$  as well as for the later discussed $a_0(980)$.

\begin{figure}[h]
    \centering
    \includegraphics[scale=0.4]{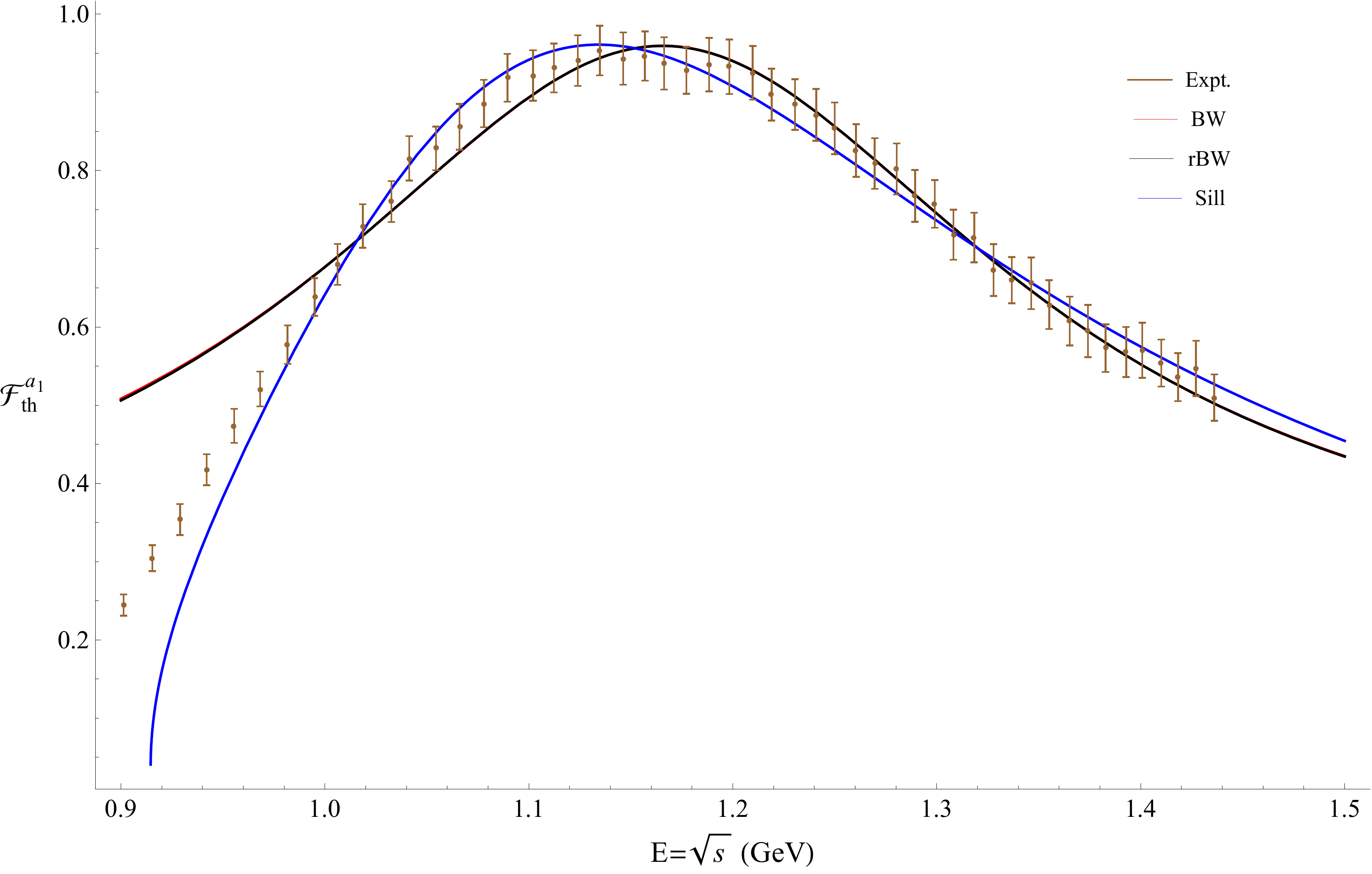}
    \caption{The spectral function for the $a_1(1260)$. Experimental data corresponds to the spectral function of the $\tau\to\pi\pi\pi\nu_\tau$ decay reported in \cite{aleph}. It is visible that the Sill fares better than the (r)BW distribution.}
    \label{a1sp}
\end{figure}

The plots of the fits of the three distributions (BW, rBW, and Sill) discussed in the text for the ALEPH data for the $a_1(1260)$ are shown in Fig. \ref{a1sp} along with the experimental data \cite{aleph}. The fitted mass and the width of the $a_1(1260)$, their error estimates, and the pole position are listed in Table \ref{a1tab}. The table shows that both the BW distributions - relativistic and non-relativistic, over-estimate the position of the peak compared to the experimental data. As indicated by the reduced $\chi^2$, the Sill distribution estimates the peak and the width of the spectral function better than the BW distributions.

The spectral function of $a_1(1260)$ poses a special problem as the peak is close to the $\rho\pi$ threshold. In the region close to the threshold, the BW and the rBW distributions overestimate the spectral function. 
As in the case of the $\rho$, the Sill can describe data fairly well even if it neglects the proper angular momentum. This fact confirms that the the inclusion of this effect, while surely relevant for a detailed description that goes beyond the simple Sill, is subleading w.r.t the threshold effect.

\begin{figure}[h]
    \centering
    \includegraphics[scale=0.5]{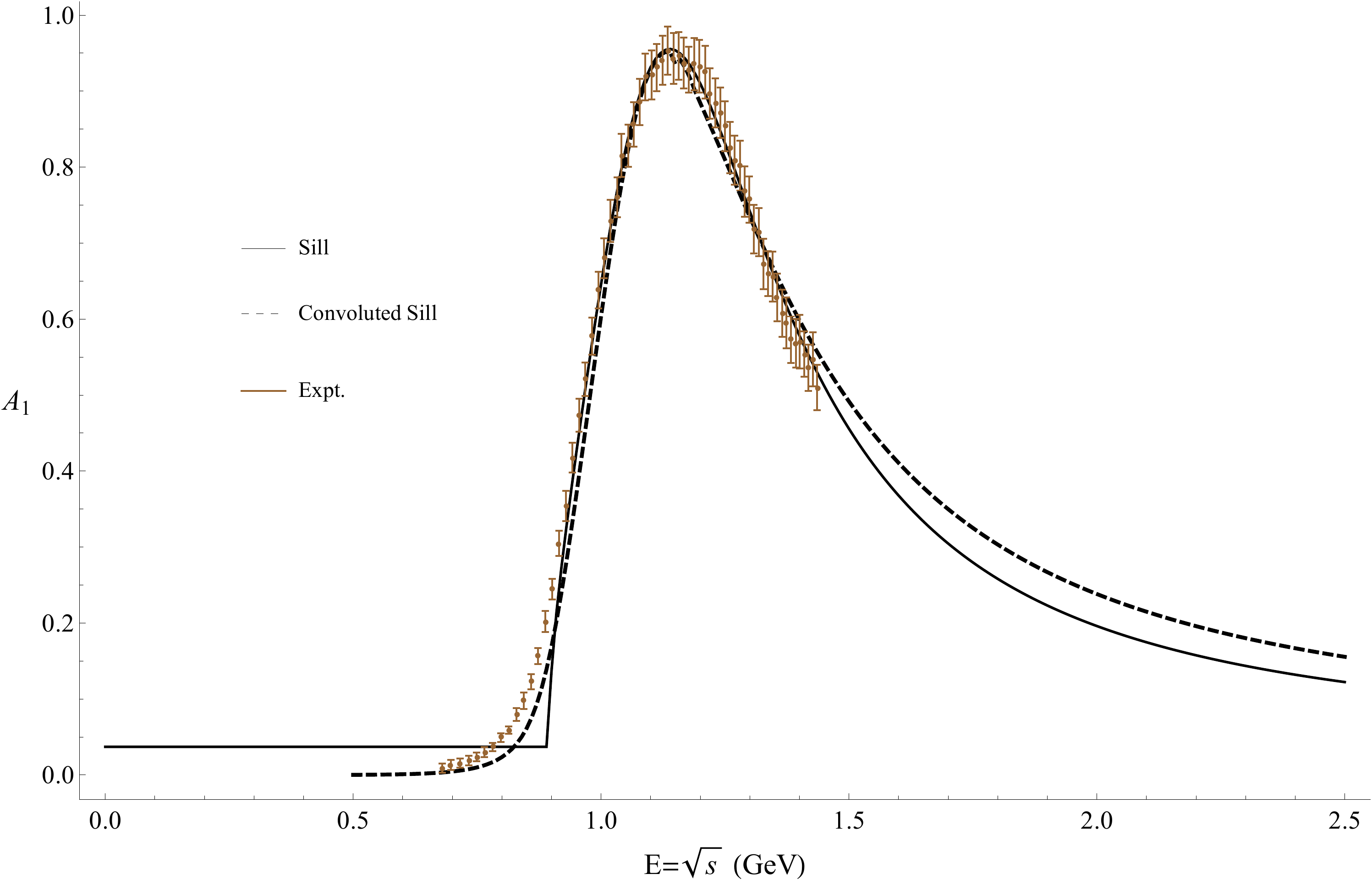}
    \caption{Comparison of the convoluted Sill and the plain Sill with the experimental data. It must be stressed that the convoluted Sill is not fitted to data (see text for details) and hence, the plain Sill fairs better in the high energy region. Yet, the convoluted Sill shows that events below the $\rho\pi$ threshold are possible.}
    \label{a1consp}
\end{figure}

The Sill distribution cannot reproduce the data close to threshold, because of the imposed threshold $m_\rho + m_\pi$. This drawback can be overcome by convoluting the spectral function of the $a_1(1260)$ with the spectral function of the $\rho(770)$, see Sec. 4.3 for details. The plot comparing the convoluted Sill and the plain Sill with the experimental data are shown in Fig. \ref{a1consp}. It is visible that the convoluted Sill can qualitatively reproduce the behavior of the spectral function near the threshold while still roughly fitting the data elsewhere. Note, however, that in this case the convoluted Sill has not been fitted to the experimental data and instead the parameters have been adjusted to fit the data points. The new values of the relevant parameters are: $g_{a_1}=1.48$ MeV, $\Gamma_{a_1}=625$ MeV, $\Gamma_\rho=130$ MeV. The rest of the parameters take the values given in Tables \ref{fittab}, \ref{rhotab} and \ref{a1tab}. In the future, it would be advisable to fit directly the convoluted Sill to data, as this would allow to describe the data points close to the $m_{\rho}+m_{\pi}$ as well as close to the peak. Here, we did not include a possible direct 3-pion decay of the $a_{1}$ meson.

\subsection{The $K^{\ast}(892)$}

\begin{figure}[h]
    \centering
    \includegraphics[scale=0.4]{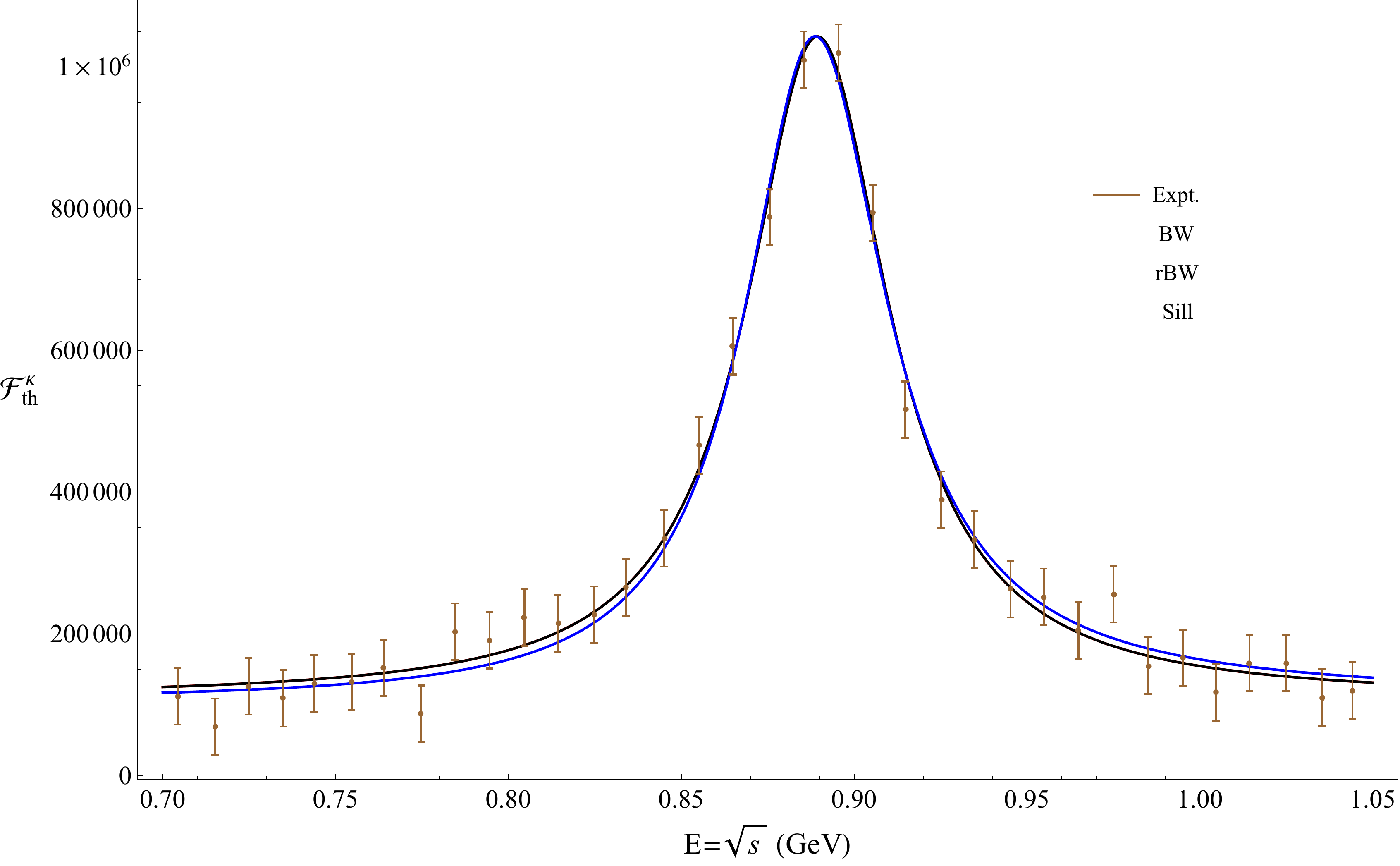}
    \caption{The spectral function for the $K^\ast(892)$. Experimental data (counts) from \cite{alicek892b}. All three functions fair similarly well. }
    \label{KSsp}
\end{figure}

The plot involving the spectral functions BW, rBW and Sill for the meson $K^\ast(892)$, fitted using the experimental data \cite{alicek892} (similar experimental data can be found in \cite{alicek892b,na61k892}) is given in Fig. \ref{KSsp}. The mass and the width used to fit the data, their error estimates, and the pole position are listed in Table \ref{KStab}. All three distributions fit the data with acceptable (and similar) accuracy. As mentioned earlier, the difference between Sill and BW and rBW distributions is negligible when $\Gamma\ll M$. 
This feature is evident in the case of $K^\ast(892)$, in which the $K\pi$ threshold is sufficiently far away from the peak to have a sizable impact on it.


\begin{table}[h]
    \centering
    \begin{tabular}{|c|c|c|c|c|}
        \hline
         Distribution & $M$ (MeV) & $\Gamma$ (MeV) & $\chi^2/\rm{d.o.f}$ & $\sqrt{ s_{pole}} (\text{MeV})$\\\hline
         Nonrelativistic BW & $889.37 \pm 0.43$ & $50.1 \pm 1.6$ & $1.78$ & $889.4 -i\, 25.0$\\
         Relativistic BW & $889.01 \pm 0.43$ & $50.1 \pm 1.6$ & $1.78$ & $890.1 -i\, 24.9$\\
         Sill & $889.06 \pm 0.43$ & $49.9 \pm 1.6$ & $2.08$ & $888.0 -i\, 25.0  $\\\hline
    \end{tabular}
    \caption{Mass and width of $K^\ast(892)$ fitted using the three distributions discussed in the text, their error estimates, and the poles (as described in the text).}
    \label{KStab}
\end{table}

\subsection{Baryonic resonances: The $\Delta(1232)$}
We now provide the example of the $\Delta(1232)$ resonance to demonstrate the applicability of the Sill distribution to the baryonic resonances. The $\Delta(1232)$ is the lightest baryonic resonances with isospin $3/2$. It has been studied extensively in the photoproduction ($N\gamma$) and $\pi$-$p$ scattering experiments. Its dominant mode of decay is into a nucleon and a pion. The PDG lists the mass and width of these resonance as $1232$ MeV and $117$ MeV respectively, and the Breit-Wigner pole at $\sqrt{s}=(1210 - i 50)$ MeV.\par
\begin{figure}
    \centering
    \includegraphics[scale=0.5]{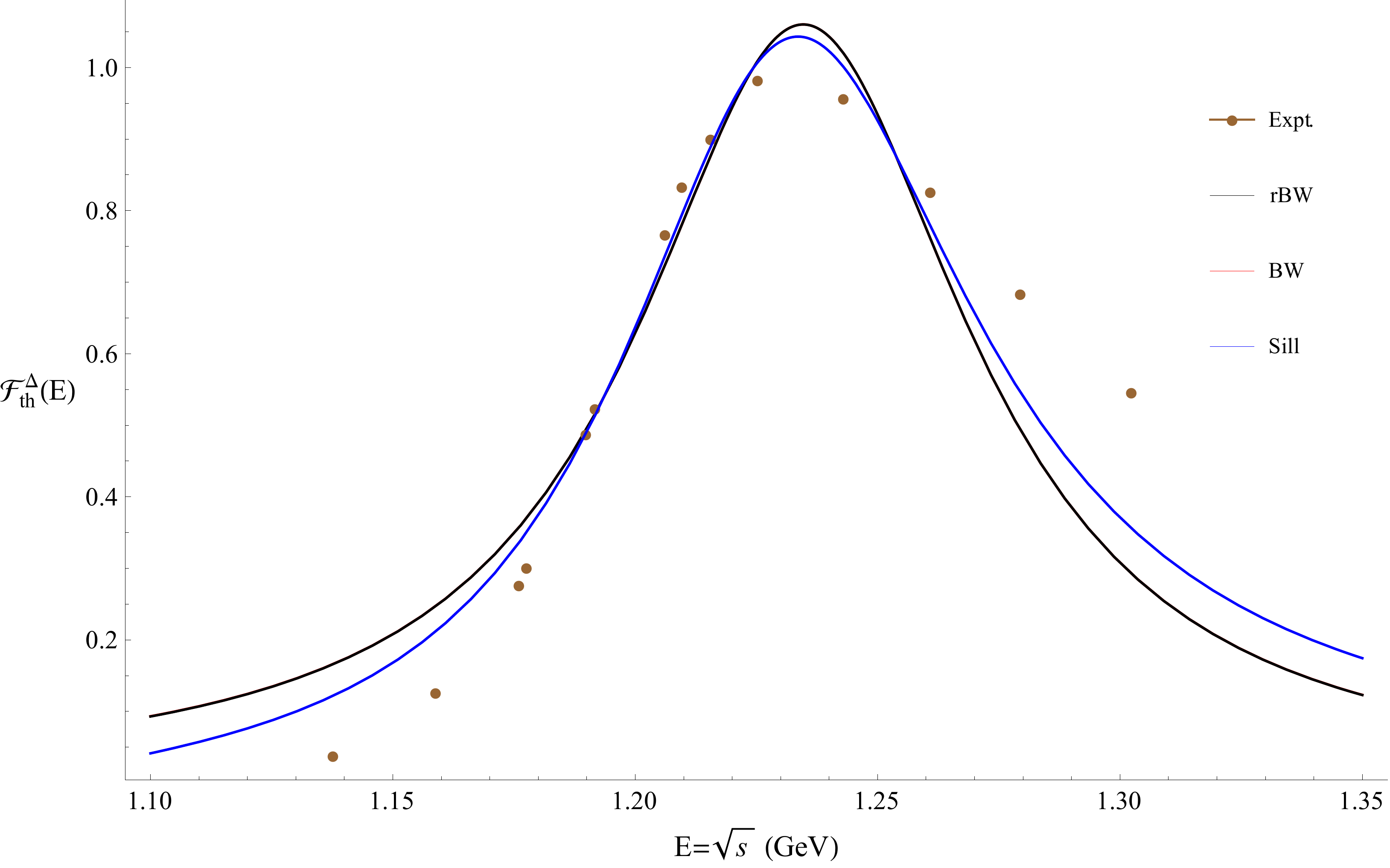}
    \caption{The spectral function for the $\Delta(1232)$. Experimental data from \cite{Haskins:1985xg}. It is visible that the Sill fairs marginally better than the (r)BW distributions.}
    \label{Delta}
\end{figure}

The BW, rBW, and Sill distributions were fitted to the spectral function of the $\Delta(1232)$ extracted from the $\pi p$ scattering cross section \cite{Haskins:1985xg}. The plots of the distributions are shown in Fig. \ref{Delta}, and the values of the parameters are listed in Table \ref{Deltatab}. The experimental data plotted in Fig. \ref{Delta} corresponds to $\sin^2(\delta(E))$, where $\delta(E)$ is the $\pi p$ scattering phase shift and this quantity is proportional to the spectral function. The error involved in this calculation has not been reported, hence we have taken the lowest value to be the estimate of the error. In addition, some remarks of caution are in order: the $\pi p$ scattering process in this energy region and angular momentum - isospin channel ($P_{33}$) gets a significant background from the $\Delta(1600)$ which has a mass of around $1510$ MeV, and width $270$ MeV for which $N\pi$ is a possible decay channel. The presence of this background can cause the spectral function to become asymmetric. We have not modelled this non-trivial background. Further, there exist non-resonant background effects which we have not taken into account (see Refs. \cite{Arndt:2006bf,Ronchen:2015vfa,Hunt:2018wqz} for some details).\par
The mass of the $\Delta(1232)$ is close to the $\pi p$ threshold. However, since the width of the $\Delta(1232)$ resonance is small, the threshold does not overlap significantly with the resonance, contrary to the case of the $a_1(1260)$. Thus, the effects of the threshold are less pronounced on the $\Delta(1232)$ resonance compared to the $a_1(1260)$. Yet, tt can be seen from the Table \ref{Deltatab} that the Sill fits better than the BW and the rBW to the data. The mass, width and the position of the pole estimated by the BW, rBW, and the Sill distributions are nearly the same. However, the Sill estimation is more accurate than the (r)BW estimate as shown by the $\chi^2$/d.o.f. Also, an inspection of the plots given in Fig. \ref{Delta} shows that although the three distributions fare identically around the peak, the Sill fits the data better closer to the threshold.

\begin{table}[h]
    \centering
    \begin{tabular}{|c|c|c|c|c|}
        \hline
         Distribution & $M$ (MeV) & $\Gamma$ (MeV) & $\chi^2/\rm{d.o.f}$ & $\sqrt{ s_{pole}} (\text{MeV})$\\\hline
         Nonrelativistic BW & $1234.6 \pm 1.3$ & $ 83.6 \pm 4.1$ & $2.92$ & $1234.7 -i\, 41.8$\\
         Relativistic BW & $1233.9 \pm 1.2$ & $ 83.7 \pm 4.2$ & $2.92$ & $1234.7 -i\, 41.8$\\
         Sill & $1236.2 \pm 1.5$ & $90.4 \pm 4.9$ & $1.53$ & $1235.4 -i\, 45.2  $\\\hline
    \end{tabular}
    \caption{Mass and width of $\Delta(1232)$ fitted using the three distributions discussed in the text, their error estimates, and the poles (as described in the text).}
    \label{Deltatab}
\end{table}

\subsection{The multi-channel sill distribution: The $f_2(1270)$}
The multi-channel Sill distribution was discussed in sec. \ref{multiSill}. Here, we demonstrate the usefulness of the Sill distribution taking the case of $f_2(1270)$ as an example.\par

The well-known resonance $f_2(1270)$ decays mostly into $\pi\pi$, $\bar{K}K$, and $4\pi$ states. 
Here, we aim to study it using the Sill in order to show how to deal with multichannel decay as well as with the convolution leading to the four-pion decay mode. This resonance is often understood as a standard quark-antiquark state belonging to a nonet of ground-state tensor mesons (yet other interpretations are possible, see Ref. \cite{CLAS:2020ngl}). 

The $\pi\pi$ channel is the dominant channel and accounts for nearly $84.2\%$ of the decays \cite{pdg}. We assume that the $4\pi$ decay proceeds through intermediate $\rho\rho$ i.e, $f_2(1270)\rightarrow \rho\rho \rightarrow 4\pi$. This is a suppressed process because the sum of the masses of the intermediate $\rho$ mesons is larger than the mass of the $f_2(1270)$. 

Out of the two observed $4\pi$ decays, the $2\pi^+2\pi^-$, which within our framework arises from the $\rho^0\rho^0$ decay, is a factor two smaller than the $\pi^+\pi^-2\pi^0$ decay, which comes from $2\rho^+\rho^-$ (this is a consequence of isospin counting). We do not distinguish between the two channels, but we note that the experimental result is consistent with our assumptions. The fraction of the $4\pi$ decay is the sum of the fractions of the two possible $4\pi$ decay channels and is $\sim 10.5\%$. With this simplification, the $4\pi$ channel becomes the second most favored channel for the decay of $f_2(1270)$.

\begin{figure}
    \centering
    \includegraphics[scale=0.4]{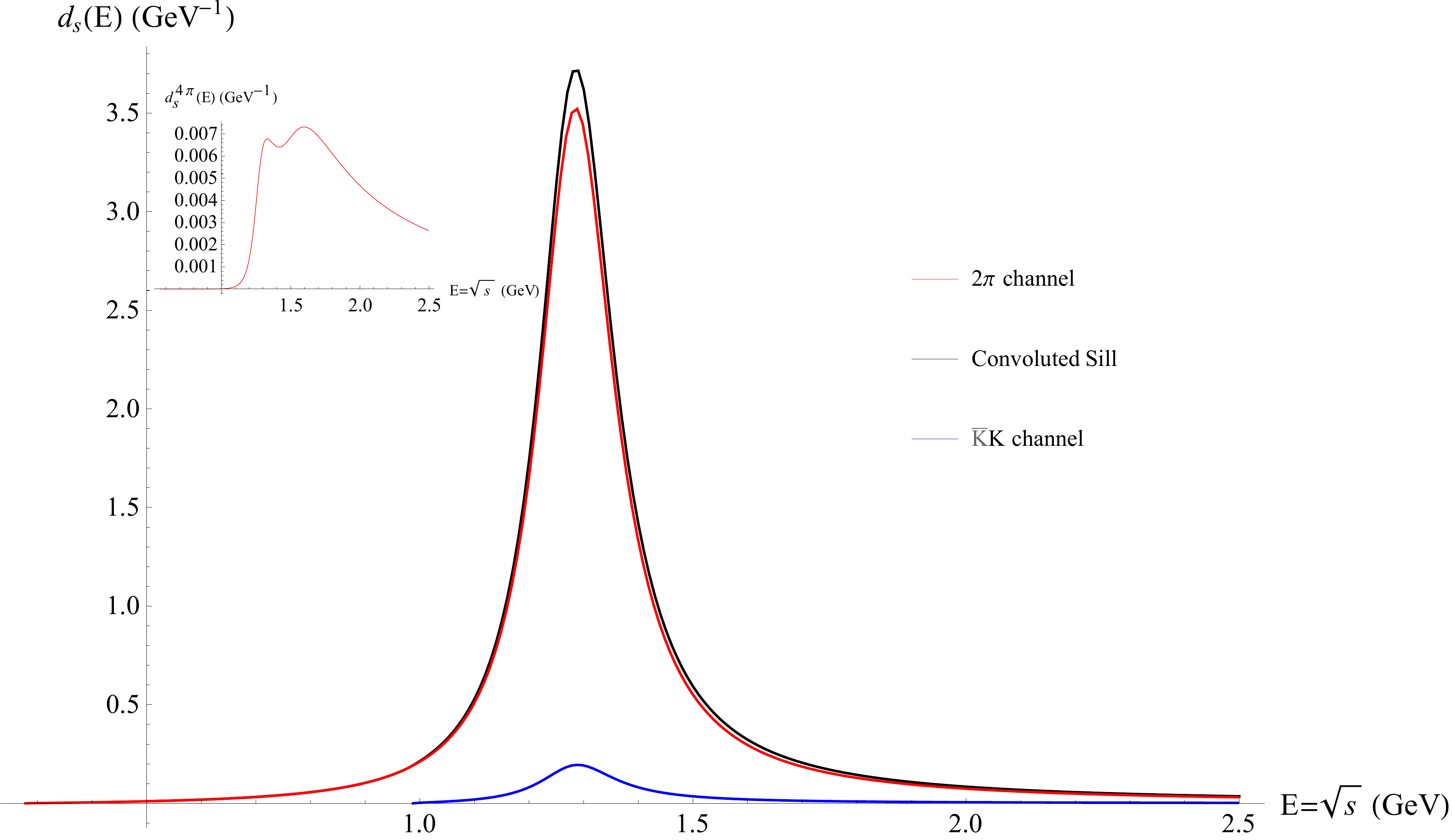}
    \caption{The convoluted Sill distribution of the $f_2(1270)$. The four channels considered are the $\pi\pi$ channel, the $\bar{K}K$ channel, and the two $4\pi$ channels. The convoluted $4\pi$ channel, whose peak is shifted to higher energies, is shown in the inset.}
    \label{f2sp}
\end{figure}

The last channel we have considered is the $\bar{K}K$ channel. This channel is also suppressed because of the high $\bar{K}K$ threshold of $\sim 980$MeV. The plot of the (convoluted) Sill distribution and the different channels is shown in Fig. \ref{f2sp}. The list of the employed parameters is given in Table \ref{f2partab}. The values of $\tilde{\Gamma_{i}}$ are calculated using Eq. (\ref{gammatilde}) with the partial decay widths calculated from the branching ratios. For the $4\pi$ channel, the $2\rho$ threshold is higher than the mass of the $f_2(1270)$. Hence, Eq. (\ref{gammatilde}) cannot be used. Instead, we use a value such that the branching ratio after convolution matches the value given in PDG \cite{pdg}. The convoluted $4\pi$ channel (inset, Fig. \ref{f2sp}) stands out because of the difference in the position of the maximum. The peak of the $\rho\rho$ mediated $4\pi$ channel appears at $\sim 1.62$ GeV, in contrast with the mass of $f_2$ which is $\sim 1.275$GeV. However, the effect of this channel is less pronounced as its contribution is small. Further, the convolution pushes the peak of the $f_2$ to a slightly higher value of $1.283$ GeV. The pole and the branching ratios extracted from the convoluted sill are listed in Table \ref{f2restab}. The values branching ratios listed here are in good agreement with the PDG.

In conclusion, the multi-channel treatment of the $f_2(1270)$ via the Sill offers an interesting, albeit simple, description which is valid in first approximation. Of course, a more detailed and correct treatment needs to make use of the correct QFT involving tensor mesons.

\begin{table}[t]
    \centering
    \begin{tabular}{cccc}
        \hline
        \hline
        \vspace*{-0.15in}&&&\\
        Mass, $M$ (MeV) & $\tilde{\Gamma}_{\pi\pi}$ (MeV) & $\tilde{\Gamma}_{\bar{K}K}$ (MeV) & $\tilde{\Gamma}_{4\pi}$ (MeV) \\\hline
        $1275.5$ & $161.1$ & $135.7$ & $21.8$\\\hline
    \end{tabular}
    \caption{Parameters of the Sill for $f_2(1270)$: the mass of $f_2(1270)$ and the (rescaled) partial widths of the three channels used in plotting Fig. \ref{f2sp}.}
    \label{f2partab}
\end{table}

\begin{table}[t]
    \centering
    \begin{tabular}{cccc}
        \hline
        \hline
        \vspace*{-0.15in}&&&\\
        $\sqrt{s_{pole}}$ (MeV) & $\text{BR}(\pi\pi)$ & $\text{BR}(\bar{K}K)$ & $\text{BR}(4\pi)$ \\\hline
        $1271.2 - i~92.7 $ & $84.8\%$ & $4.65\%$ & $10.6\%$\\\hline
    \end{tabular}
    \caption{The pole and the branching ratios of the three decay channels of $f_2(1270)$ extracted from the Sill distribution.}
    \label{f2restab}
\end{table}

\begin{figure}
    \centering
    \includegraphics[scale=0.4]{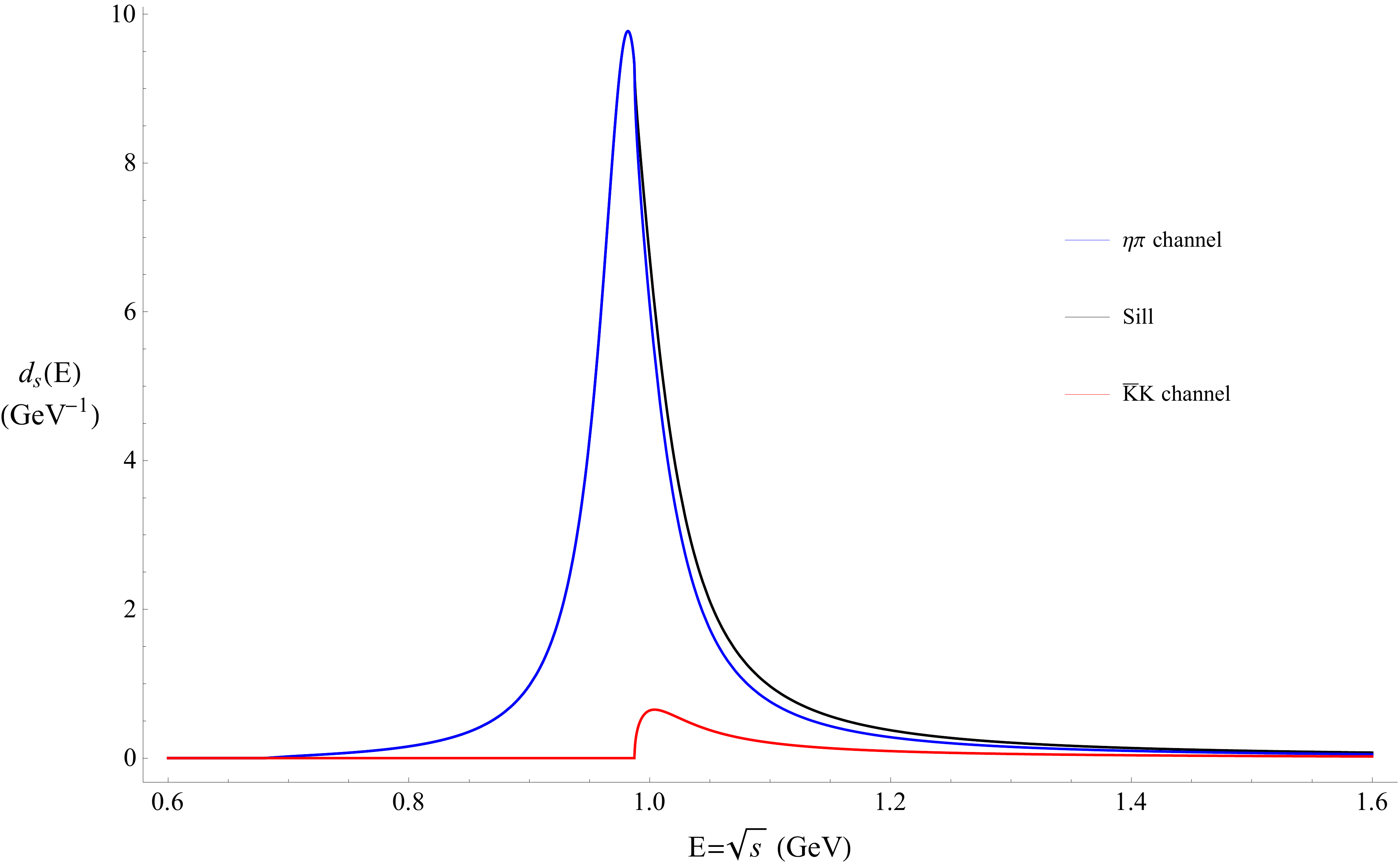}
    \caption{The Sill distribution of the $a_0(980)$ and the $\eta\pi$ and $\bar{K}K$ channels. The non-BW form due to the $KK$ threshold is evident. }
    \label{a0sp}
\end{figure}

\subsection{Further examples and limitations}

In this subsection, we present some additional examples which show the limitations of the Sill. 

\subsubsection{The $a_0(980)$}

The resonance $a_0(980)$ is peculiar, since it was used to introduce for the first time the well-known Flatt\'{e} distribution, which is generalized by the Sill in this work. Moreover, the kaon-kaon mode of this resonance opens just at the mass of this resonance, thus making it an interesting case for a multichannel study. 
On the other hand, care is needed, since the resonance $a_0(980)$ has been studied in a variety of works as a exotic meson candidate (e.g. in the form of a molecular state or a companion pole) ({\it e.g,} \cite{Dudek:2016cru,Jaffe:1976ig,Weinstein:1990gu,Baru:2003qq,Pelaez:2004xp}, for a review on the status of exotic hadrons, see \cite{Klempt:2007cp}).

We thus proceed as follows. First, we discuss the $a_0(980)$ as a single normalized state, ignoring its nontrivial nature, and later on we include modification/extension due to its non-conventional origin. 

The $a_0(980)$ decays predominantly into the $\eta\pi$ and the $\bar{K}K$ channels. As discussed in Ref. \cite{Bugg:2008ig}, the $\eta\pi$ channel is the dominant decay channel for $a_0(980)$. The $\bar{K}K$ channel is special here, as the mass of the resonance lies very close to the $\bar{K}K$ threshold.  
The branching ratios of the decay $B^0 \rightarrow \psi a_0(980) \rightarrow \bar{K}K$ is very much sensitive to the mass of the $a_0(980)$ \cite{rui}.

The  multi-channel Sill is capable of explaining  some of the peculiar features of the shape of its spectral function. As numerical inputs, we use the branching ratios and the width at half-maximum given in the PDG \cite{pdg} to estimate the parameters listed in Table \ref{a0partab} employed for plotting Fig. \ref{a0sp}.  
While this study should be only regarded only as heuristic, it represents a useful demonstration of a channel opening right at the threshold. In fact, this threshold effect is visible in the spectral function as well as in both the decay channels at $\sqrt{s}=2m_K$.
For the parameters given in Tabe \ref{a0partab}, the pole occurs at $\sqrt{s_{pole}}=973 - i36$ MeV.


\begin{table}[h]
    \centering
    \begin{tabular}{ccc}
        \hline
        \hline
        \vspace*{-0.15in}&&\\
        Mass, $M$ (MeV) & $\tilde{\Gamma}_{\eta\pi}$ (MeV) & $\tilde{\Gamma}_{\bar{K}K}$ (MeV)\\\hline
        $980.0$ &  $91.2$ & $44.0$\\\hline
    \end{tabular}
    \caption{Parameters of the Sill for  $a_0(980)$: the mass and the rescaled partial widths of the $\eta\pi$ and $\bar{K}K$ channels used in plotting Fig. \ref{a0sp}.}
    \label{a0partab}
\end{table}

Yet, as mentioned above, the $a_0(980)$ is nowadays regarded as a non-conventional dynamically generated mesonic state. 
In this context, the state associated with $a_0(980)$ does not need to be normalized to unity (as naively done in Fig. \ref{a0sp}), as approaches in which this state is interpreted a kaon-antikaon bound state show \cite{Baru:2003qq}. Moreover, a similar conclusion is obtained in Refs. \cite{boglione,Wolkanowski:2015lsa,vanBeveren:1986ea}, in which $a_0(980)$ is described as a companion pole of the predominantly quark-antiquark state $a_0(1450)$. 
In both pictures, the normalization of $a_0(980)$ amounts to a number $Z<1$.\par
In this respect, the question is how to modify the Sill in order to take into account the more complicated nature of both resonances $a_0(980)$ and $a_0(1450)$.
We thus write down a normalized two-Sill as a simple model for two-state system in which one is a companion pole of the other as: 
\begin{align}
    d_S^{a_0}(E) &= Z~\!d_S^{low}(E) + (1-Z)~\!d_S^{high}(E) \text{ ,}
    \label{EffSill}
\end{align}
where the superscripts $low$ and $high$ refer to the $a_0(980)$ and the $a_0(1450)$ respectively and where $Z$ (with $0\leq Z\leq1$) is a weight factor. Thus, the two-Sill is a simple parametrization aimed to capture some of the features of rather complex phenomenon that involve hadronic loops. More specifically, in the large-$N_c$ limit one expects a unique quark-antiquark state $\left\vert \bar{q}q\right\rangle$, whose mass is about 1.3 GeV and corresponds (roughly) with $a_0(1450)$.  Then, the interaction of the original seed state with other mesons induces a mixing that adds new components to the Fock space that are proportional to $\left\vert \bar{K}K\right\rangle $ and $\left\vert
\pi\eta\right\rangle$. Thus, the original state is a mixture of both quark-antiquark and meson-meson component. This mechanism is quite general and takes place also for standard resonances, such as the already mentioned $\rho$-meson. In the specific case of the $a_0$-system, the mixing can be strong enough to
generate an additional pole that is identified  to $a_{0}(980)$ (the $\kappa-K_0^*(1430)$ system is also similar, as discussed in the next subsection), which, as a consequence, can be seen as predominantly a meson-meson state. 

\begin{figure}
    \centering
    \includegraphics[scale=0.4]{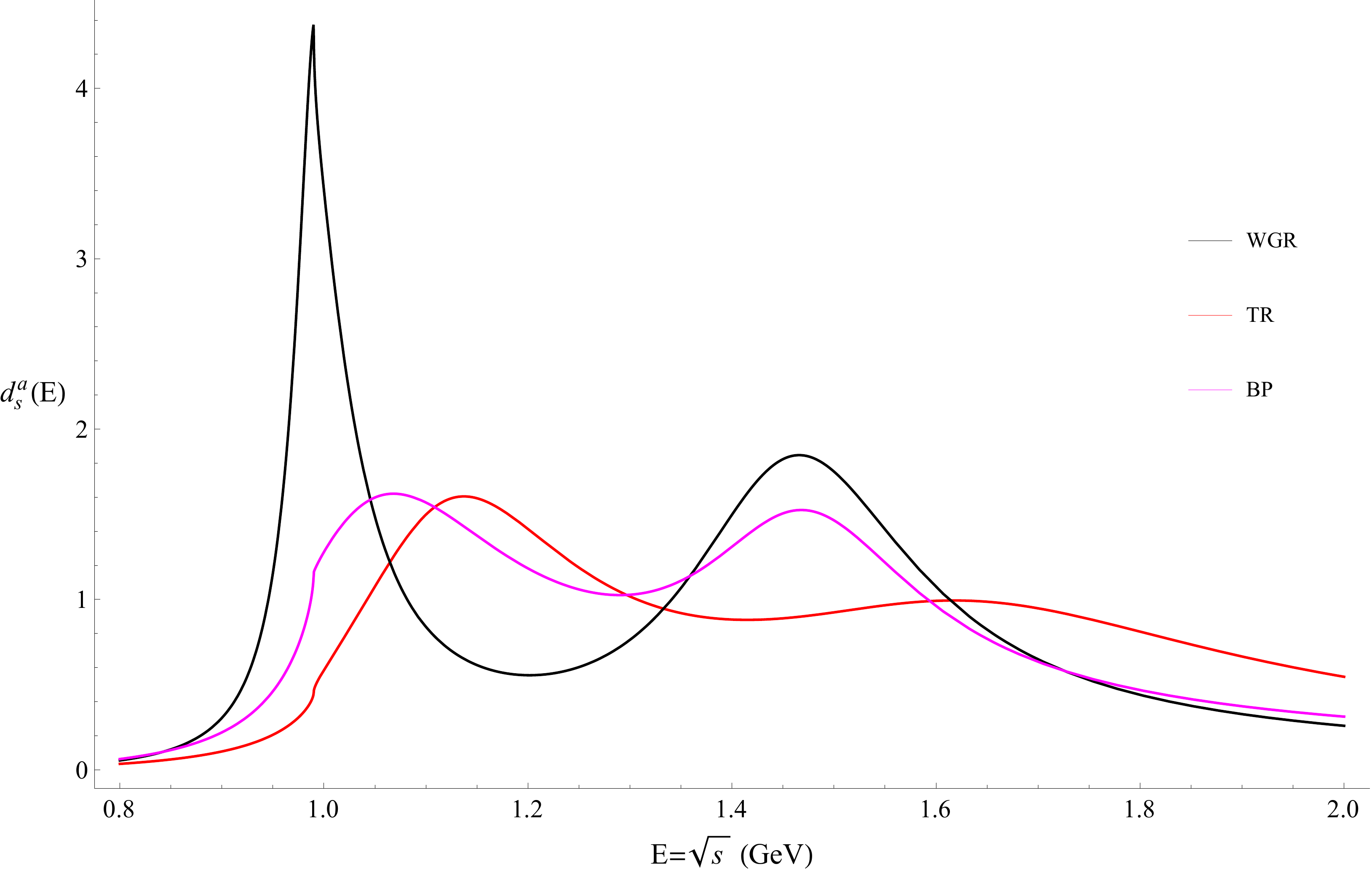}
    \caption{The Sill distribution of the $a_0(980)$ and the $a_0(1450)$. For the details on the legend, see text.}
    \label{a0full}
\end{figure}

In Fig. \ref{a0full}, we plot the two-Sill distribution for the poles given by Tornqvist and Roos (TR) \cite{Tornqvist:1995ay}, Boglione and Pennington (BP) \cite{boglione}, and Wolkanowski {\it et al} (WGR) \cite{Wolkanowski:2015lsa}, in which the corresponding loop calculations were performed.

The values of these poles for $a_0(980)$ are:
\begin{align}
    \sqrt{s_{pole}} &= 1094 - i145\text{ MeV (TR)}\nonumber\\
    \sqrt{s_{pole}} &= 970 - i45\text{ MeV (WGR)}\nonumber
\end{align}
and for the $a_0(1450)$:
\begin{align}
    \sqrt{s_{pole}} &= 1592 - i284\text{ MeV (TR)}\nonumber\\
    \sqrt{s_{pole}} &= 1456 - i134\text{ MeV (WGR)}\nonumber
\end{align}
Note, BP provide the pole mass and pole width as $M_{pole}=1082$ MeV and $\Gamma_{pole}=309$ MeV. For the $a_0(1450)$, we have used the BW mass and width given in the PDG \cite{pdg}. 

The constant $Z$ is taken to be $0.6$, $0.5$, and $0.7$ for the TR, WGR, and the BP curves respectively, in such a way to mimic those corresponding calculations (see the plots for all 3 cases in Ref. \cite{Wolkanowski:2015lsa}). 

The two-Sill distributions in Fig. \ref{a0full} are actually pretty similar to those of TR, BP, and WGR, showing that a qualitative agreement is offered by our simple approach. 
Clearly, it must be stressed again that the two-Sill does not represent a substitute of a full loop calculation, but it shows that the presented modification of the Sill offers a potentially useful way to study, in the first approximation, some non-conventional resonances.

As a final remark, we underline that the companion pole scenario presented above is only one framework among the possibilities to understand the dynamically generated states. 
Other relevant approaches exist, for instance the already mentioned molecular picture for the $a_0(980)$ of Ref. \cite{Baru:2003qq}, which is based on the compositeness condition \cite{Weinberg:1965zz,Hayashi:1967bjx}. 
Quite interestingly, in Ref. \cite{Baru:2003qq} the integral over the spectral function between $2m_K-0.05$ GeV and $2m_K+0.05$ GeV ranges between 0.2-0.5. For comparison, a standard BW spectral function with a width of 50 MeV would deliver $(2/\pi) \arctan(2) \sim 0.7$. The fact that obtained value (0.2-0.5) is smaller than 0.7 was interpreted as a sign of an important molecular component in this state (an even smaller value is found for the $f_0(980)$). A similar integral for the three curves of Fig. 9 delivers the values 0.05 (TR), 0.1 (BP), and 0.25 (WGR), the last of which is consistent with \cite{Baru:2003qq}.
Surely, this result is built-in by the form of Eq. (\ref{EffSill}), which is inspired by previous works on the subject. 
Nevertheless, as a general remark, the lack of normalization of a certain resonance around its peak can be seen as a hint of its non-standard nature.

\subsubsection{The $\kappa$ and $K_0^\ast(1430)$}

\begin{figure}[h]
    \centering
    \includegraphics[scale=0.4]{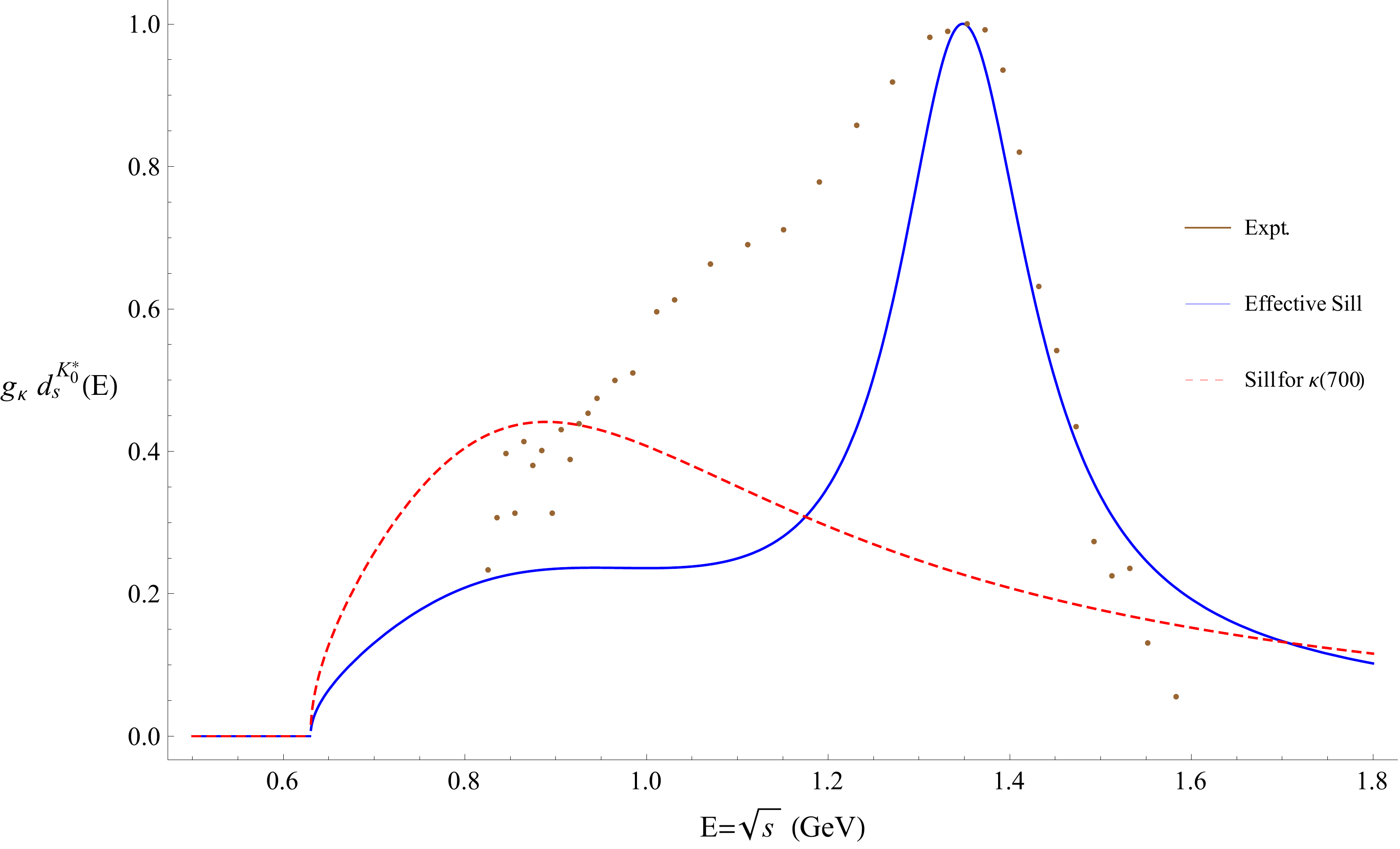}
    \caption{The Sill distribution of the $\kappa$ and the $K_0^\ast(1430)$. Experimental data from \cite{Aston:1987ir}.}
    \label{kappa}
\end{figure}

The next question is if the description mentioned above for the $a_0(980)$ can be applied also to the resonance $\kappa$, which does not even show a peak in the cross-section data. 

Indeed, also here the normalization of $\kappa$ is not a requirement, since this state is also (predominantly) not quark-antiquark mesons, see e.g. Refs. \cite{wolkak,Pelaez:2016klv}.

As an exercise, we try to apply the two-Sill to this case as well. We thus consider $\kappa$ as the companion pole of $K_0^\ast(1430)$ and write down: 
\begin{align}
    d_S^{K_0^\ast}(E) &= Z~\!d_s^{low}(E) + (1-Z)~\!d_s^{high}(E) \text{ ,}
    \label{EffSillkappa}
\end{align}
where in this case low refers to  $\kappa$ and high to $K_0^\ast(1430)$.

The pole of the $\kappa$ is listed by the PDG as $\sqrt{s_{pole}}=(630-730)-i(260-340)$ MeV. 
We adjust the mass and the width of the $d_s^{low}(E)$ to get the pole at $\sqrt{s_{pole}}\approx700-i295$ MeV. 
The PDG does not list a value for the pole of $K_0^\ast(1430)$. We thus adjusted the mass and width of the $K_0^\ast(1430)$ to obtain the peak seen in the data, which leads to a pole at $\sqrt{s_{pole}}=1345-i88$ MeV. The weight factor $Z$ takes the value $0.5$. 

The masses and widths so obtained were used to calculate the effective distribution similar to in Eq. \ref{EffSill} and the resultant curve is plotted in Fig. \ref{kappa}, along with the data from the $\pi K$ scattering experiment \cite{Aston:1987ir}. The plotted experimental values correspond to $\sin^2(\delta_{\pi K}(E))$, where $\delta_{\pi K}(E)$ is the $\pi K$ scattering phase shift. Also plotted in the Fig. \ref{kappa} is the spectral function of the $\kappa$.

It is visible that the two-Sill is peaked at about 1.3 GeV and the light $\kappa$ represents only an enhancement, as more advanced model show. 
Yet, it is not possible to get a qualitative agreement if we impose that the pole of the light $\kappa$ agrees with the PDG range. (A better agreement could be obtained at the price of a pole at higher values, but this is in conflict with many works on the subject). 

We thus encounter a clear quantitative limitation of the double-Sill for this specific case. This is somewhat expected: the light $\kappa$ is a pole that is caused by pion-kaon loop effects, and the (double-)Sill is not capable of reproducing the details of this system. 

For what concerns the $f_0(500)$, the Sill that corresponds to the pole quoted by the PDG has a similar form as that of the light $\kappa$ reported in Fig. 10, but is even more elongated. This is in agreement with the non-conventional nature of this state. A more detailed study of this sector would necessitate even more than two resonances, but the situation is at least as complex as the light $\kappa$: one does not expect to get a correct description of data for such a non-perturbative system using the Sill.

\begin{table}[t]
    \centering
    \begin{tabular}{|c|c|c|c|c|c|c|}
    \hline
        Meson & Mass (MeV) & Width (MeV) & Channel & $\Delta m$ (MeV) & $BR$ ($\%$) & $k$ (MeV) \\\hline
        $\pi_1(1600)$ & $1661$ & $240$ & $b_1\pi$ & $296$ & $seen$ & $357$\\
        $\psi(3770)$ & $3774$ & $27$ & $D\bar{D}$ & $34$ & $93^{+8}_{-9}$ & $287$\\
        $\psi(4040)$ & $4039$ & $80$ & $D^*\bar{D}^*$ & $25$ & $seen$ & $226$\\
        $\chi_{c1}(4140)$ & $4147$ & $22^{+8}_{-7}$ & $J/\psi\phi$ & $30$ & $seen$ & $217$\\
        $\Upsilon(4S)$ & $10580$ & $20.5$ & $\bar{B}B$ & $22$ & $>96$ & $326$\\\hline
    \end{tabular}
    \caption{Illustrative list of some mesonic resonances near the respective thresholds. $\Delta m$ is the difference between the masses of the initial and final states.}
    \label{thrmes}
\end{table}

\begin{table}[h]
    \centering
    \begin{tabular}{|c|c|c|c|c|c|c|}
        \hline
        Baryon & Mass (MeV) & Width (MeV) & Channel & $\Delta m$ (MeV) & $BR$ ($\%$) & $k$ (MeV) \\\hline
        \multirow{3}{*}{$N(1535)$} & $1515$-$1545$ & $125-175$ & $N\pi$ & $440-470$ & $32-52$ & $464$\\
        & $1515-1545$ & $125-175$ & $N\pi$ & $440-470$ & $32-52$ & $464$\\
        & $1515-1545$ & $125-175$ & $N\eta$ & $27-57$ & $30-55$ & $176$\\
        \hline
        $\Delta(1600)$ & $1500-1640$ & $200-300$ & $\Delta\pi$ & $133-273$ & $73-83$ & $276$\\\hline
        \multirow{2}{*}{$\Lambda(1600)$} & $1570-1630$ & $150-250$ & $N\bar{K}$ & $136-196$ & $15-30$ & $343$\\
        & $1570-1630$ & $150-250$ & $\Sigma\pi$ & $246-306$ & $10-60$ & $338$\\\hline
        $N(1720)$ & $1680-1750$ & $150-400$ & $N\omega$ & $(-42)-28$ & $12-40$ & $-1$\\
        \hline
    \end{tabular}
    \caption{Illustrative list of some baryonic resonances near the respective thresholds. $\Delta m$ is the difference between the masses of the initial and final states. The masses mentioned in the table are the Breit-Wigner masses.}
    \label{thrbar}
\end{table}
\subsection{Applicability of the Sill}

In this subsection, we discuss under which conditions one may expect the proposed Sill distribution to be -in first approximation- useful.

An important feature of the Sill distribution is its (simple but consistent) treatment of the thresholds. 
As the various examples in Sec. 5 have shown, the Sill works better than (r)BW  when dealing with the states/resonances lying close to the threshold. In particular, denoting by $\Delta m$ as the distance to a certain threshold, we expect some advantage of using the Sill when $\Gamma \sim \Delta m$. This condition is indeed fulfilled for the $a_1$-meson with $\Gamma \sim 300 $ MeV and $\Delta m \sim 300$ MeV and for the $\Delta$-baryon with $\Gamma = 117$ MeV and $\Delta m = 157$ MeV. Even in the case of the $\rho$-meson, for which $\Gamma \sim 150 $ MeV is three times smaller than $\Delta m \sim 500$ MeV (thus $\Gamma/\Delta m \sim 1/3$) the fit with the Sill is favored. On the other hand, for $K^{*}(892)$ (for which $\Gamma \sim 50$ MeV and $\Delta m \sim 250$ MeV, resulting in a value $\Gamma/\Delta m \sim 1/5$) the Sill and the (r)BW deliver very similar results, thus in this case the use of the Sill does not provide additional advantages. 
Out of these considerations, it seems that a ratio $\Gamma/\Delta m \gtrsim  1/3$ is required for the Sill to be advantageous. For values smaller than this ratio we do not expect the Sill to be better than (r)BW. 

Interestingly, states that have masses lying close to significant thresholds appear quite often. 
Some of such states have been listed in Table \ref{thrmes} and Table \ref{thrbar} along with their masses, decay widths, prominent decay channels whose threshold is close to the mass of the parent, the distance to the threshold, branching ratios (when known), and the 3-momentum carried by the decay products. 

In the meson sector, notable examples are the four heavy mesons $\psi(3770)$, $\psi(4040)$, $\chi_{c1}(4140)$, and $\Upsilon(4S)$. For instance, the $\psi(3770)$ and the $\Upsilon(4S)$ are very close to the $D\bar{D}$ and the $B\bar{B}$ threshold respectively \cite{pdg}. As one can see from the Table \ref{thrmes}, the difference between the masses of the parent and the decay products is less than or very close to the decay widths of the parent states. In this Table, we include also the hybrid candidate  $\pi_1(1600)$ (note, as an hybrid, this state is not dynamically generated, but a confined state that survives in the large-$N_c$ limit). For this resonance, the lattice calculations show that the $b_1\pi$ channel is the dominant channel \cite{Woss:2020ayi}.

Also among baryons there are resonances which have one of the dominant modes of decay as sub-threshold. One such example is the $N(1720)$. The $N(1720)$ is a $3/2^+$ baryon with the two dominant decay channels: $N\pi\pi$ and $N\omega$. Of these, the $N\omega$ decay channel is sub-threshold with the mass difference between $-42$ MeV and $28$ MeV. Other resonances, like the $\Delta(1600)$, the $\Lambda(1600)$, and the $N(1535)$ lie very close to the thresholds of their dominant decay channels, as displayed in Table \ref{thrbar}.

The closeness of these states to their respective thresholds is expected to alter their line shapes, making the Sill as a valid alternative to the conventional relativistic Breit-Wigner distribution functions.
Of course, the Sill is a quite general parametrization with some specific interesting features, that however does not represent a substitute to more complete studies that take into account the properties related to the specific quantum numbers as well a more advanced inclusion of loop effects of a given state.

The resonances listed in the Tables can be interpreted  as conventional quark-antiquark mesons or three-quark baryons (with the exception of the already mentioned hybrid $\pi_{1}(1600)$ resonance).
In case of dynamically generated states, care is needed, as we discuss in detail in Sec. 5.6. One may still use the Sill -e.g. in the framework of the companion pole scenario- but the results can be at best qualitative. 

Summarizing, the Tables \ref{thrmes} and \ref{thrbar} should be seen as illustrative examples (beyond Sec. 5) that show that threshold effects are pretty common. Yet, for those states, that are analogous to the resonances discussed in Sec. 5,  we may expect the Sill to be useful.  In the future, whenever a certain resonance sufficiently close to a threshold is found, together with the usual (r)BW, we think that a fitting via the Sill could be advisable. 

As a final remark, we mention also that the Sill can be employed as toy model. The fact that the normalization of the spectral function is fulfilled also for broad states close to the thresholds, make it an interesting tool for testing its properties that appear in certain hadronic models, for instance, when calculating decays by taking into account finite width effects or as an intermediate state in three-body decay.

\section{Conclusions}

In this work, after a brief recall of the non-relativistic  and relativistic Breit-Wigner functions and their justifications as certain limiting cases, we have presented a relativistic and properly normalized distribution -called Sill distribution- which models the presence of energy threshold(s). It can be in principle applied to any unstable state, but the realm of hadrons 
seems the best place to use such a spectral function, since hadrons are often quite broad and various thresholds are present. Indeed, the form of the Sill is inspired by the way the kaon-kaon decay mode was included 
in the so-called Flatt\'{e} distribution for the resonance $a_0(980)$. The form of the self-energy is applied to any decay channel (even if the masses of the decay products are different), regardless of the particular angular momenta involved in the decay. 

The determination of the pole is also straightforward within the Sill distribution. Moreover, the extension to a the multichannel case and to a convolution suitable for decay chains have been also developed. 

A direct comparison with a scalar QFT shows under which conditions the Sill function emerges. Moreover, a direct application of the Sill to the well-known resonances $\rho$ and $a_1(1260)$ shows that, despite its simplicity, it works pretty well. This is especially evident for the latter case, due to the fact that $a_1(1260)$ is quite broad.
A similar, even if less pronounced, improvement has been achieved for the baryonic state $\Delta(1232)$, showing that the Sill can be also applied to baryonic resonances.
On the other hand, for the relatively narrow resonance such as $K^*(892)$, the Sill distribution fairs as well as the standard BW and rBW functions. 
We have also shown a direct implementation of the multichannel and convoluted sill for the well established resonances $f_2(1270)$.

Next, we have also discussed the example of non-conventional mesons, such as the resonance $a_0(980)$. A possible simple description is to extend the Sill to describe the $a_0(980)$-$a_0(1450)$ system. Yet, in doing so we have encountered also limitations of the Sill, that -as expected- become evident also in the case of the very broad and unconventional light $\kappa$ meson. 


In conclusion, we have presented a simple and general distribution that can be applied for various resonances, including broad ones affected by threshold effect(s). Since the expected effort of using such a distribution in fits to data w.r.t. the standard relativistic Breit-Wigner seems comparable and in consideration of the relative simplicity and theoretical foundation of behind it, which includes the normalization necessary for a probability distribution, we regard the Sill as an additional useful tool to describe some resonances. 

\bigskip

\textbf{Acknowledgments: } The authors thank M. Rybczynski for useful discussions. F. G. and V. S. acknowledge financial support through the Polish National Science Centre (NCN) via the OPUS project 2019/33/B/ST2/00613.

\appendix

\section{Alternative relativistic Breit-Wigner distribution}\label{rBWE}
\begin{figure}[h]
    \centering
    \includegraphics[scale=0.4]{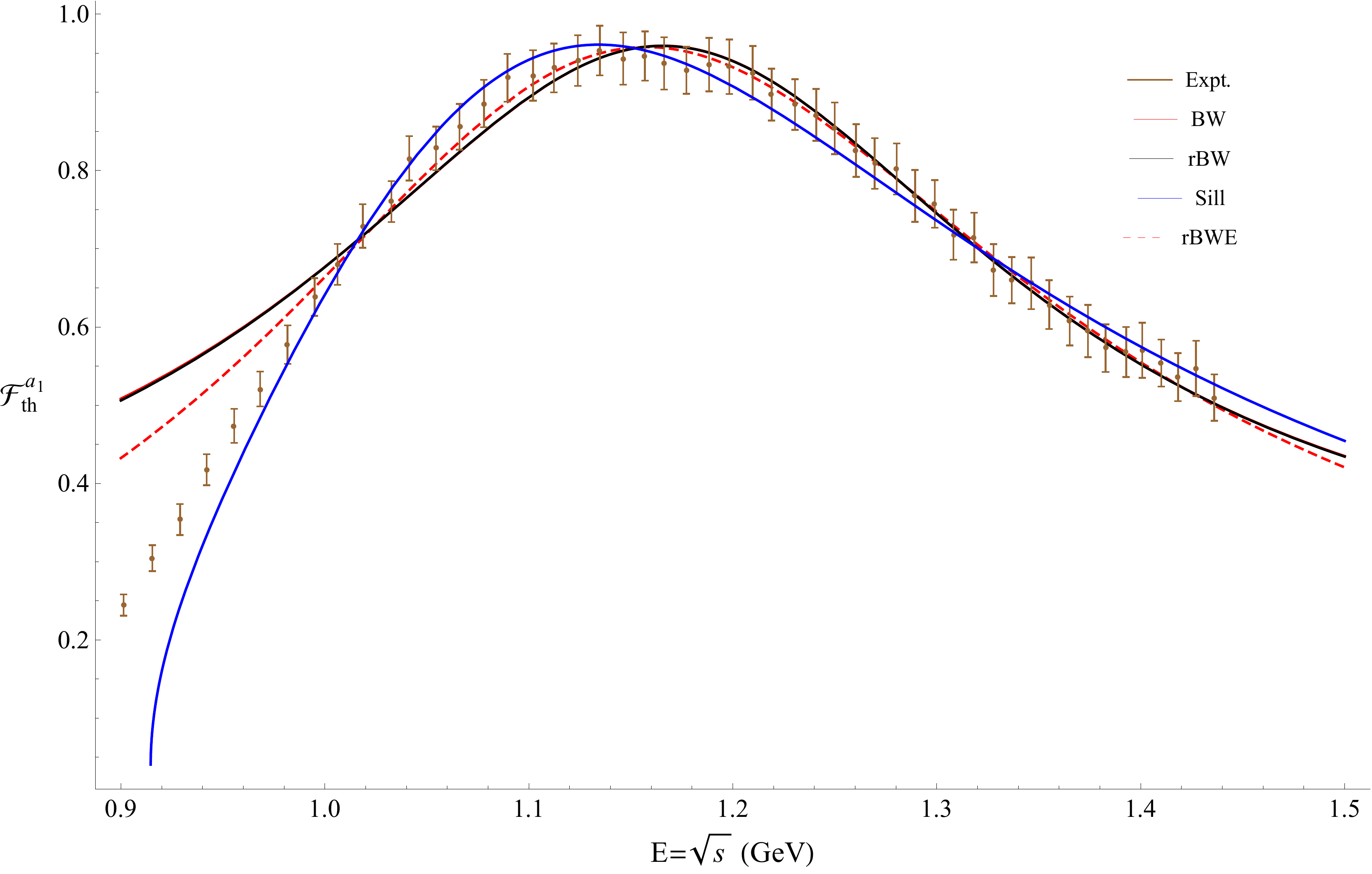}
    \caption{The spectral function for the $a_1(1260)$. The rBWE and rBW have been labeled as rBWE and rBW respectively.}
    \label{rBWEp}
\end{figure}

In this appendix we discuss a relativistic by keeping in which we keep the energy dependence in the numerator.  The corresponding function, denoted as rBWE, reads: 
\begin{equation}
    d_{S}^{\text{rBWE}}(E)=\frac{2E}{\pi}\frac{E\Gamma}{(E^{2}-M^{2}%
)^{2}+(E\Gamma)^{2}}\theta(E)\text{ .}\label{rBWEeq}
\end{equation}
This expression differs from the one given in Eq. (\ref{rBWM}) in the numerator and the denominator. This alters the form of the distribution slightly. For the purpose of demonstration, we fit Eq. (\ref{rBWEeq}) to the spectral function of the $a_1(1260)$. The $\chi^2$ fit returns us the following parameters: $g_{a_1}=646.9$ MeV, $M=1157.7\pm1.2$ MeV, $\Gamma= 475.0\pm 12.7$ MeV, and $b_{a_1} = 0.091$, with a $\chi^2/$d.o.f of $4.25$. The pole appears at $\sqrt{s_{pole}}=1181-i 233$ MeV. These values are to be compared with the values given in Table \ref{a1tab}. We see that, the mass of the $a_1(1260)$ estimated by rBWE is significantly larger than that estimated by rBW and smaller than the value estimated by the Sill. The real part of the pole is larger than that estimated by rBW and the Sill, and the imaginary part lies in between the values estimated by the rBW and the Sill. We display in Fig. \ref{rBWEp} the plots of the fits of rBWE along with all the other forms the distribution functions discussed in the text and the experimental data from ALEPH \cite{aleph}.

From the plots and the parameters, we see that the two versions of the rBW distributions provide very similar fits to the data, and estimated values of the mass and width are close to those estimated by the nonrelativistic BW. The fits themselves are better than that of the nonrelativistic BW as evidenced by the $\chi^2/$d.o.f value but worse than the fit of the Sill distribution.\par

\begin{figure}[h]
    \centering
    \includegraphics[scale=0.4]{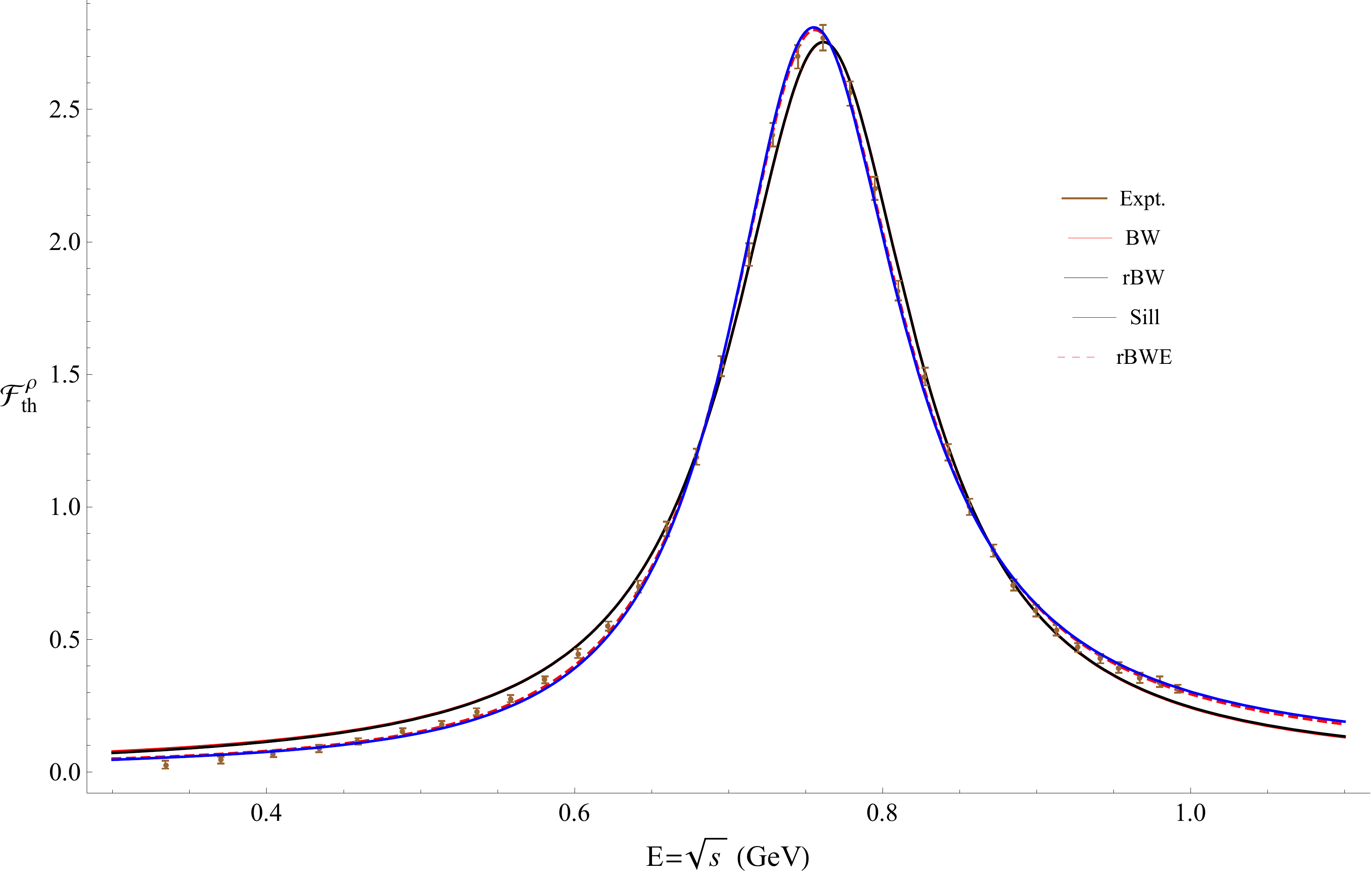}
    \caption{The spectral function for the $\rho(770)$. The rBWE and rBW have been labeled as rBWE and rBW respectively.}
    \label{rBWEprho}
\end{figure}

On the other hand, the spectral function of the $\rho(770)$, as estimated by the 4 distributions appear similar, as shown in Fig. \ref{rBWEprho}. However, the mass, width, and the pole positions estimated by the rBWE are signifcantly different from that estimated by the rBW or the Sill. The rBWE estimates for these quantities are: $M=756.1\pm0.30$ MeV, $\Gamma=139.0\pm1.2$ MeV, and $\sqrt{s_{pole}}=759.2-i69.2$ MeV, with a $\chi^2/$d.o.f of $9.42$. The mass and width are closer to the values estimated by the Sill, but the position of the pole is closer to the value estimated by rBW. This estimate is as accurate as the rBW estimate as per the value of $\chi^2/$d.o.f. Since the $\rho(770)$ peak is far away from the $2\pi$ threshold, one cannot expect the Sill to show a large deviation from the rBW or the rBWE. However, according to the statistics, the Sill fits the resonance better than the rBW (or the rBWE).

In general, we may notice that there is not a unique rBW form. 

\section{Hesse matrix and error estimation}\label{hesse}
The errors in the parameters have been calculated using the Hesse matrix. Here, we provide a brief description of the Hesse matrix and error estimation. Consider a set of data that is expected to be described by some theoretical functions $f_i$, each of them depending on the $N$ parameters $\{x_1,~x_2,~\ldots,~x_N \}$. The goodness of the fit can be measured using the $\chi^2$ defined as:
\begin{equation}
\chi^2 (x_1,~x_2,~\ldots,~x_N ) = \sum_{i=1}^{m}\left( \frac{f_i(x_1,~x_2,~\ldots,~x_N )-f_{exp,i}}{\delta f_{exp,i}} \right)^2 \text{ ,}
\end{equation}
where the $f_{exp,i}$ are the measured values of the function $f_i$ and the $\delta f_{exp,i}$ are the error in the measurements. The parameters can be estimated by demanding that the value of $\chi^2$ is minimum. The Hesse matrix ($H$) is defined as the matrix of the second derivatives of the $\chi^2$, 
\begin{equation}
    H_{ij} = \left(\frac{1}{2} \frac{\partial^2 \chi^2(x_1,~x_2,~\ldots,~x_N )}{\partial x_i \partial x_j}\right)_P
\end{equation}
where $P$ represents the values of the parameters for which $\chi^2$ is at its minimum. With this definition of the Hesse matrix, the errors in the parameters are given by
\begin{equation}
    \delta x_i = \sqrt{H^{-1}_{ii}} \text{ .}
\end{equation}

\end{document}